\documentclass[%
reprint,
nofootinbib,
mathtools,
amsmath,
amssymb,
prd,
longbibliography,
superscriptaddress]{revtex4-1}
\usepackage[T1]{fontenc}
\usepackage[latin9]{inputenc}
\usepackage[dvipsnames]{xcolor}
\usepackage{amsmath}
\usepackage{float}
\usepackage{array} % Required for the 'w' column specifier
\usepackage{amssymb}
\usepackage{graphicx}
\usepackage{esint}
\usepackage{soul}
\usepackage{multirow}
\usepackage{enumitem} % needed for enumerate
\usepackage{pgfplots}

\usepackage{comment}
\usepackage{amsfonts}
\usepackage{babel}
\usepackage[pdftex,breaklinks,colorlinks,
linkcolor=Blue,
citecolor=teal,
anchorcolor=red,
urlcolor=cyan]{hyperref}
\usepackage[capitalise]{cleveref}

\usepackage{caption}
\usepackage{subcaption}
\usepackage{siunitx}

\graphicspath{{Figures/}{TikzPlots/}}

\newcommand{\PB}[1]{{\color{Blue}{PB: #1}}}

\begin{document}
	\title{Not All Resonances Are Created Equal: Prioritizing Tidal Resonances in EMRIs}
	
	\author{B\'eatrice Bonga}
	\email{bbonga@science.ru.nl}
	\affiliation{Institute for Mathematics, Astrophysics and Particle Physics, Radboud
		University, 6525 AJ Nijmegen, The Netherlands}

    \author{Patrick Bourg}
    \email{patrick.bourg@ru.nl}
\affiliation{Institute for Mathematics, Astrophysics and Particle Physics, Radboud
		University, 6525 AJ Nijmegen, The Netherlands}
    \author{Bram ten  Brink}
\affiliation{Institute for Mathematics, Astrophysics and Particle Physics, Radboud University, 6525 AJ Nijmegen, The Netherlands}
    \author{H.A. (Bart) Peters}
	\email{\textcolor{black}{Current affiliation: Faculty of Electrical Engineering, Mathematics and Computer Science, Delft University of Technology, The Netherlands}}
\affiliation{Institute for Mathematics, Astrophysics and Particle Physics, Radboud University, 6525 AJ Nijmegen, The Netherlands}

	\date{\today}
	%\pacs{xxxx}
	\begin{abstract}
	Tidal perturbations from nearby compact objects can drive resonant ``kicks'' in extreme-mass-ratio inspirals (EMRIs), imprinting potentially detectable phase shifts in LISA-band gravitational waveforms. We provide a systematic survey of these tidal resonances across EMRI parameter space in the weak-tide, three-body hierarchy, treating the perturber as stationary on the resonance timescale. For generic Kerr orbits we compute (i) resonance contours in $(p,e,x)$ that locate where each $(n,k,m)$ resonance is encountered, (ii) the associated resonance duration, and (iii) the jump amplitudes in the angular momentum $L_z$ and the Carter constant $Q$ (with $\Delta E$ vanishing by time-translation symmetry in the stationary model). Combining jump amplitudes with resonance durations, we construct a practical ranking of the resonances most relevant for waveform modeling, while emphasizing that even modest resonances may still be important through their influence on the phase at which subsequent resonances are entered. All contour and jump data are publicly available. 
	\end{abstract}
	\maketitle
 
\section{Introduction}
Extreme-mass-ratio inspirals (EMRIs) are a key target for LISA, given that they are unique probes of the spacetime structure around the massive central black hole (BH) \cite{Barack:2003fp}. They will allow for extraordinarily precise tests of general relativity, since we will observe their intricate relativistic orbits over some $\sim 10^5$ orbits instead of the typical $O(10)$ orbits in comparable-mass binaries observed by LIGO-Virgo-KAGRA. EMRIs are typically modeled as clean, isolated 2-body systems for which the astrophysical environment is often neglected. 
However, recent progress in the field has shown that environments are important and without properly accounting for them, we risk systematic biases in parameter estimation and may fail to perform the promised high-precision tests of the nature of black holes. Environmental effects include accretion disks \cite{KocsisYunesLoeb2011,Duque:2024mfw,Duque:2025yfm,HegadeKR:2025dur,HegadeKR:2025rpr,Dyson:2026ddd}, dark matter  \cite{Duque:2023seg,Hannuksela:2019vip} and axion clouds \cite{Zhang:2018kib,Dyson:2025dlj}. Another key environmental effect is nearby stars or black holes.

During most of the EMRI inspiral, the tidal field of nearby objects can be neglected, provided that these objects are at reasonable distances. However, at certain times during the evolution, nearby stellar-mass objects can induce tidal resonances. These resonances leave an observable imprint on the gravitational waveform \cite{Bonga:2019ycj}. Without accounting for these resonant effects in our model, the utility of EMRIs as precision probes may be impeded. We may misattribute these resonance effects to deviations from General Relativity or, worse, may not even be able to detect the EMRI. Conversely, if we can correctly model such tidal resonances, we unlock valuable information about the population of dark objects in the galactic core, otherwise difficult to access \cite{Gupta:2021cno,Gupta:2022fbe}. 

Observational evidence has accumulated over the past years indicating the reality of such tidal perturbers with the observations of QPEs possibly describing the interaction of a stellar object with an accretion disk of a central massive BH \cite{Miniutti:2019fqr,Franchini:2023bou,Chakraborty:2024tzd,Kejriwal:2024bna,Zhou:2025udg,Liu:2026tvy,Chen:2026fdv} and the discovery of new faint stars near SgrA* (such as S301 with a periapsis at only $\sim 260 M_*$  with $M_*$ the mass of SgrA*\cite{S301}). While EMRI rates with or without perturbers remain highly uncertain \cite{Babak:2017tow,Pan:2021oob}, this highlights the importance of correctly incorporating tidal resonances in EMRI models.

The urgency of this work is driven by practical considerations for LISA data analysis. Our goal is to incorporate tidal resonances into the Fast EMRI Waveforms (FEW) framework. FEW is an open-source code for the rapid generation of accurate EMRI waveforms \cite{Chua:2020stf,Katz:2021yft,Speri:2023jte,Chapman-Bird:2025xtd} and is incorporated into the LISA Distributed Data Processing Centre (DDPC) pipeline. Because FEW is highly modular, it is well suited to include tidal-resonance effects, provided we can determine (i) when a resonance occurs and (ii) the size of its effect, which we refer to as the \emph{resonance jump}.

In this paper, we systematically map the resonances induced by a tidal perturber across the EMRI parameter space, treating the perturber as stationary on the resonance timescale. We compute the resonance contours that determine where resonances occur, and we evaluate the associated jump sizes.

Although many resonances may be encountered during a typical inspiral, not all are excited strongly enough to leave observable imprints on the waveform. Excitation depends sensitively on the phase with which the system enters the resonance. The jump size scales as (resonance strength)$\times$(resonance duration)$\times e^{i\,\mathrm{phase}}$. The resonance strength is directly proportional to the strength of the tidal field and depends non-trivially on the resonance numbers and orbital configuration. The resonance duration is determined by the amount of radiation emitted in the form of gravitational waves, and is therefore  generally longer for resonances occurring at larger semi-latus rectum (i.e., farther from the central black hole). The phase can take any value in $[0,2\pi)$ and is extremely sensitive to the initial conditions with which the EMRI enters the resonance.

All resonance contours and the associated jump data are made publicly available \cite{GitHubRepo}. Using these results, we provide a practical hierarchy of the most important tidal resonances for EMRIs. Two caveats are worth emphasizing. First, even a large instantaneous jump may have limited impact on the waveform if it occurs very close to the separatrix, as it will affect only the final few orbits. Second, because the jump depends on the waveform phase, even modest resonances may influence subsequent resonances by shifting the phase at which they are entered. How important this latter effect is, is currently being investigated.

This paper is organized as follows. In Sec.~\ref{sec:set-up} we introduce the system and modeling assumptions, and we define the resonance condition and jump. Section~\ref{sec:contours} presents the resonance contours across parameter space. Section~\ref{sec:duration} evaluates the resonance duration. In Sec.~\ref{sec:jumpamps}, we present the resonance strengths and combine these ingredients to rank resonances by expected importance. We conclude in Sec.~\ref{sec:discussion} with implications for LISA data analysis and directions for future work.

Throughout this paper, we use geometrized units with $c = G = 1$, where $c$ is the speed of light and $G$ is the gravitational constant. 
%When converting geometric units to astronomical units (AU), we use $1\,\mathrm{AU} \approx 25\,M_\star$, where $M_\star$ is the mass of Sagittarius~A$^\star$. 
Any dimensionful quantity, such as the black-hole spin $a$ and semi-latus rectum $p$, is normalized by appropriate factors of the central black-hole mass $M$.

\section{Formulation}
\label{sec:set-up}
In this section we summarize the physical set-up and the approximations used in the remainder of the paper. Our treatment follows Refs.~\cite{Gupta:2021cno,Bart-master-thesis}; additional technical details of the jump calculation are collected in App.~\ref{app:jumps}.

\subsection{System and scale hierarchy}
\begin{figure}
\includegraphics[width=0.45\textwidth]{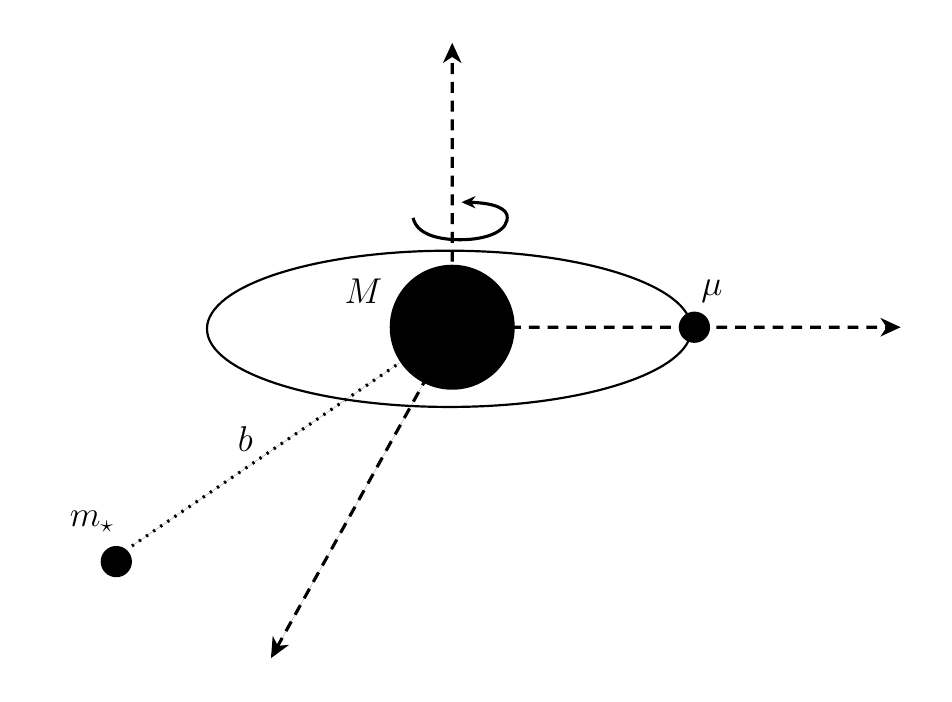}
  \caption{An EMRI system, consisting of a massive central black hole of mass $M$, an close inspiral object of mass $\mu \ll M$, and a third body far away at distance $b$, acting as the tidal perturber with mass $m_\star$.}
\label{fig:set-up}
\end{figure}

We consider an EMRI consisting of a central massive black hole of mass $M$ and an inspiraling compact object of mass $\mu$, subject to the tidal field of a third body (the \emph{(tidal) perturber}) of mass $m_\star$ at a large separation $b$ from the central black hole; see Fig.~\ref{fig:set-up}. We assume the mass hierarchy $M \gg \mu, m_\star$ and the weak-tide regime $M/b \ll 1$.

%When quoting explicit numbers we use $M = 4\times 10^6\,M_\odot$ (similar to the mass of Sagittarius~A$^\star$), $\mu = m_\star = 30\,M_\odot$ (corresponding roughly to the peak of mass distribution inferred from LIGO-Virgo-KAGRA observations), and $b = 10\,\mathrm{AU}$ (for which $M/b \sim 4\times 10^{-3}$). 

\subsection{Forced-geodesic description}
In the limit $\mu\to 0$ and in the absence of the perturber, the small body follows a geodesic of the Kerr spacetime. To incorporate the external tide we adopt the standard forced-geodesic picture in which the inspiraling object is treated as an accelerated point particle moving in the Kerr background \cite{Poisson:2011nh,Sago:2008id,Spallicci:2014bja}. The acceleration induced by the external perturbation can be written as
\begin{align}\label{eq:acc}
a^\alpha & = -\frac{1}{2} (g^{\alpha\beta}_{\rm Kerr}+u^\alpha u^\beta)(2h_{\beta \lambda;\rho}-h_{\lambda \rho; \beta}) u^\lambda u^\rho\;,
\end{align}
where $u^\alpha$ is the four-velocity of the small body. Here $h_{\mu\nu}$ denotes the tidal metric perturbation sourced by the distant perturber (constructed by matching a suitable post-Newtonian tidal metric to a deformed black-hole geometry).

For sufficiently weak tides, the direct, non-resonant impact of Eq.~\eqref{eq:acc} on the long-term inspiral is negligible \cite{Barausse2014}. The effects we target here arise instead when the system crosses a resonance \cite{Bonga:2019ycj,Gupta:2021cno,Gupta:2022fbe}.

\subsection{Resonance condition, duration, and ``jump''}
A resonance occurs when the Kerr orbital frequencies (defined with respect to Boyer--Lindquist time) become commensurate,
\begin{equation}
    \omega_{nkm} := n\,\omega_r + k\,\omega_\theta + m\,\omega_\phi = 0,
    \label{eq:resonance-condition}
\end{equation}
for integers $n,k,m$. As in the self-force resonance problem \cite{PhysRevLett.109.071102}, the resonance is transient because radiation reaction slowly evolves the orbital frequencies. The associated resonance timescale is
\begin{equation}
    T^{\rm res}_{nkm} = \sqrt{\frac{4\pi}{\tfrac{\mu}{M}|\Gamma_{nkm}|}},
    \label{eq:ResDuration}
\end{equation}
with
\begin{equation}
\Gamma_{nkm} := n\,\dot{\omega}_r + k\,\dot{\omega}_\theta + m\,\dot{\omega}_\phi\;,
\end{equation}
where overdots denote derivatives with respect to Boyer-Lindquist time evaluated as the orbit crosses the resonance surface.

Across a resonance, the orbit experiences a discrete change (a \emph{jump}) in its orbital constants of motion, which we denote schematically by $\Delta J_{i}$. Our goal in this paper is to determine where resonances occur and to quantify these jump amplitudes; the explicit expressions used for $\Delta J_i$ are given in App.~\ref{app:jumps}.

Self-force resonances are restricted to $m=0$ by axisymmetry \cite{Flanagan:2010cd}, whereas the tidal field generically excites $m\neq 0$ and thus a larger family of resonances.

\subsection{Stationary-perturber approximation}
The evolution of the EMRI frequencies is dominated by radiation reaction and therefore scales with the mass ratio $\mu/M$, implying $T^{\rm res}_{nkm} \propto \sqrt{M/\mu}$. For typical EMRIs this corresponds to $\mathcal{O}(10^2\text{--}10^3)$ orbital cycles, depending on the orbital parameters and the labels $(n,k,m)$. Over such timescales the perturber would also evolve; in this work we assume it is sufficiently distant so that its motion can be neglected during the resonance (``stationary tide''). This is expected to be accurate for short-duration resonances (e.g., those encountered in the strong field), while for longer-duration resonances the perturber's orbital phase can change appreciably. An extension including the perturber's orbital evolution will be presented elsewhere \cite{inpreparation}; fully dynamical three-body regimes have been explored numerically in Refs.~\cite{Silva:2022blb,Silva:2025lkl}.

\subsection{Tidal-field model for $h_{\mu\nu}$}
In principle, for generic EMRIs with dimensionless spin $0\le a < 1$, one should compute the tidal perturbation using Kerr perturbation theory (e.g., Ref.~\cite{Yunes:2005ve}\footnote{Beware that an overall factor of two is missing in $h_{\alpha\beta}$ in~\cite{Yunes:2005ve}; see footnote 17 in~\cite{LeTiec:2020bos} for details.}). In practice, we find that (i) tidal resonances are strongest for slowly rotating black holes \cite{Gupta:2021cno,Gupta:2022fbe}, and (ii) spin effects in the tidal metric perturbation enter at higher post-Newtonian order while substantially increasing computational cost. For clarity and speed (a factor of $\sim 20$--$50$ for the stationary resonances considered here), we therefore compute $h_{\mu\nu}$ using Schwarzschild perturbation theory in lightcone gauge \cite{poisson2015tidal}, while still computing geodesic motion and orbital frequencies in a Kerr background. 
This choice also naturally includes the $m=0$ modes, which were excluded from the analysis in \cite{Yunes:2005ve}. (Other descriptions of a tidally perturbed Kerr black hole exists that do include these modes; see, e.g., the outgoing radiation gauge construction in Ref.~\cite{Cocco:2026lkr}.)

We verified that, for the stationary $(3,0,-2)$ resonance, the resulting resonance strengths differ by $\lesssim 0.1\%$ across our parameter space. For highly eccentric orbits, the discrepancy can reach $\sim 1\%$. This comparison is performed at the level of the (gauge-invariant) jump amplitudes, since a direct comparison of $h_{\mu\nu}$ is gauge dependent.

Following Refs.~\cite{Bonga:2019ycj,Gupta:2021cno,Gupta:2022fbe}, we model the black hole's tidal environment using only the leading-order (Newtonian) contribution, which is then matched onto a deformed black-hole spacetime. At this order, the tidal field is fully characterized by the electric-type quadrupole moment. Higher-multipole and post-Newtonian corrections are suppressed by additional powers of $M/b$ and/or by the perturber's orbital velocity \cite{poisson2015tidal}.
%%%%%%%%%%%%%%%%%%%%%%%%%%%%%%%%%%%%%%%%%%
%%%%%%%%%%%%%%%%%%%%%%%%%%%%%%%%%%%%%%%%%%
\section{Resonance contours}
\label{sec:contours}
%%%%%%%%%%%%%%%%%%%%%%%%%%%%%%%%%%%%%%%%%%
%%%%%%%%%%%%%%%%%%%%%%%%%%%%%%%%%%%%%%%%%%
 %(calculation of contours, separatrix, banding at low a, 
%lowering of contours at high a and I, n=5,6 contour behavior, 
%missing n=2 contours, little amount of n=1 contours, frequency plots,
% resonances that dive partially under separatrix, arising difficulties with waveform modeling
%  dynamic case: influence of s and consequences, (maybe showing resonance finder behavior with  double contours))\\ \\
\label{subsec:staticprogcontours}

%\paragraph{Key points.}
%\begin{itemize}
  %\item We consider all low-order resonances with $-4 \leq n \leq 4$, $-4 \leq k \leq 4$ and $-2 \leq m \leq 2$, with the quadrupolar tidal perturbation restricting $|m| \leq 2$.
  %\item Symmetry and selection rules reduce 405 formal combinations to 110 potential resonances, then to 24 resonances that lie within the LISA band and have nontrivial jumps.
  %\item Prograde and retrograde orbits are treated separately, with prograde resonances using $k<0$ and/or $m<0$ and retrograde using $k<0$ and/or $m>0$.
  %\item The prograde contour shapes are largely similar in eccentricity, with $n=1$ resonances tracking the separatrix and $n=2$ resonances absent over the relevant parameter space.
  %\item Since $\omega_r$ is always smaller than $\omega_\theta$ and $\omega_\phi$, only sets with $|n| > |k + m|$ can resonate at low $p$; for $n=1$, this effectively requires $k + m = 0$.
  %\item The weighted frequency sum $\omega_{nkm}(p)$ is not injective, rising sharply through zero at low $p$ and decreasing at larger $p$, converging to $(n+k+m)\omega_{\text{kep}}$.
  %\item Higher-$n$ resonances with the same $k+m$ cluster near the separatrix and are expected to be less impactful than the $n=1$ resonances.
%\end{itemize}

Resonance surfaces are found by solving Eq.~\eqref{eq:resonance-condition} at a given resonance number $(n,k,m)$.
In principle, infinitely many triplets $(n,k,m)$ give rise to a corresponding solution. For most of this paper, we will consider all low-order resonances with $-4 \leq n \leq 4$, $-4 \leq k \leq 4$ and $-2 \leq m \leq 2$, for a total of $9 \times 9 \times 5 = 405$ possible combinations of ($n, k, m$).~\footnote{The perturbation is modeled by the leading order quadrupolar deformation of the central massive BH, which naturally restricts $|m| \leq 2$.} 
It is expected that resonances with higher ($n$, $k$, $m$) lead to significantly smaller jumps in the orbital constants. We have explicitly verified this for a few resonances with $n=5$ and $n=6$. In particular, comparing the jumps in the angular momentum and Carter constant for resonance contours $(n,-1,-1)$, with $n=3,4,5$ for $a=e=x=0.5$, we find that indeed for higher $n$ the jump values decrease: $||(\langle dL_z/dt \rangle, \langle dQ/dt \rangle)_{(3,-1,-1)}|| = (12.2, 134.9)$, while $||(\langle dL_z/dt \rangle, \langle dQ/dt \rangle)_{(4,-1,-1)}|| = (3.9, 39.8)$, and $||(\langle dL_z/dt \rangle, \langle dQ/dt \rangle)_{(5,-1,-1)}|| = (2.2, 22.2)$. Furthermore, since the resonance contour for $(5,-1,-1)$ also lies much closer to the separatrix than its $n=3, 4$ counterparts, see Fig.~\ref{fig:extracontours}, its impact on the waveform is expected to be much smaller. Similarly, the jump amplitudes associated with a higher-$n$ resonance contour are typically smaller than those in neighboring lower-$n$ contours with the highest jumps. For example, the $(6,-3,1)$ resonance lies in the $n=3$ resonance band, specifically between the $(3,-1,-1)$ and $(3,-2,0)$ resonance contours, and its jump amplitudes are $||(\langle dL_z/dt \rangle, \langle dQ/dt \rangle)_{(6,-3,1)}|| = (0.5, 8.6)$. For comparison, the two largest jumps with $n=3$ are the $(3,-1,-1)$ (jumps given above) and $(3,0,-2)$ where $||(\langle dL_z/dt \rangle, \langle dQ/dt \rangle)_{(3,0,-2)}|| = (11.3, 42.0)$.
For completeness, $||(\langle dL_z/dt \rangle, \langle dQ/dt \rangle)_{(3,-2,0)}|| = (0.026, 47.9)$, $||(\langle dL_z/dt \rangle, \langle dQ/dt \rangle)_{(3,-3,1)}|| = (3.5, 63.3)$, and
$||(\langle dL_z/dt \rangle, \langle dQ/dt \rangle)_{(3,-4,2)}|| = (0.9, 10.0)$.

\begin{figure}
  \centering
  \includegraphics[width=0.47\textwidth]{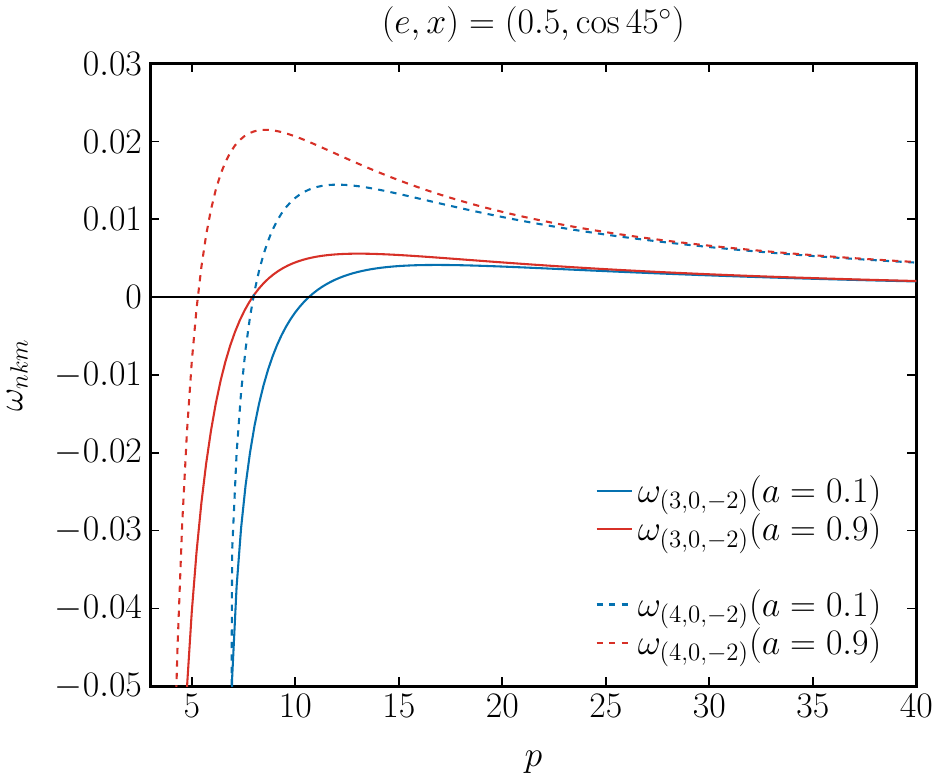}
  \caption{The weighted sum of the frequencies, $\omega_{nkm}$, for two resonance frequencies at two different spins, at fixed eccentricity $e = 0.5$ and inclination $x = \cos 45^\circ$. The qualitative behavior of $\omega_{nkm}$ is the same for retrograde orbits.}
  \label{fig:omegankm}
\end{figure}

The weighted sum of the frequencies $\omega_{nkm}$ as a function of the semi-latus rectum $p$ is shown in Fig.~\ref{fig:omegankm} for two representative resonances, i.e., the $(3, -4, 2)$ and $(3, 0, -2)$ resonances.
It can be seen that the $\omega_{nkm}$ of these two resonances have a similar dependence on $p$. This shape of the $\omega_{nkm}$ seems to occur for all resonances with $n \leq 4$.
One important feature to note is that the dependence of $\omega_{nkm}$ on $p$ is not injective. The curves increase very fast from negative values to positive values at low $p$. Then, the curves slowly decrease with increasing $p$.
At large $p$, the weighted sum of the frequencies converge to 
\begin{equation}
  (n+k\pm m)\omega_{\text{kep}},
  \label{eq:Keplerian}
\end{equation}
where $\omega_{\text{kep}}$ is the Keplerian frequency $\sqrt{\frac{M + \mu}{p^3}}$. The $\pm$ correspond to pro- and retrograde orbits, respectively. As a result, it is expected that there is no more than one solution to the stationary resonance condition.

Most of the $405$ combinations of $(n,k,m)$ we consider do not yield a resonance with a jump, so these numbers are filtered out as much as possible using the following set of general selection rules:
\begin{enumerate}
  \item Since ($n$, $k$, $m$) and ($-n$, $-k$, $-m$) represent the same resonance, we will only consider resonances where the first nonzero number is non-negative.
  \item At least two of the triplets $(n,k,m)$ have to be non-zero.
  \item It has been shown that the tidally induced jumps in $L_z$ and $Q$ are invariant under reflections of the orbit in the equatorial plane \cite{Bart-master-thesis}. 
  This means that the jumps in $L_z$ and $Q$ are zero unless $k + m =$ even, so only resonances with $k + m =$ even will be considered.
  \item If there exists a set of resonance numbers ($n$, $k$, $m$), there is a resonance with numbers ($jn$, $jk$, $jm$) at the exact same location for any integer $j$ with $|j| > 1$.
  It is expected that generically these resonances with ($jn$, $jk$, $jm$) are less important since they have higher numbers, so they will be ignored. However, there is one exception to this: if a lower resonance is discarded due to point 3, it could be relevant here. For instance, we include $(4,0,-2)$ even though it is a scaled version from $(2,0,-1)$,  because $(2,0,-1)$ has no jump as $2-1=1$ is odd, however $4-2=2$ is even.
  %  \item When $n=0$, we found that the resonance contours are extremely close to the separatrix. We therefore remove these cases too, as they are irrelevant for EMRI modeling. This removes an additional 12 potential resonances.
\end{enumerate}
The application of the above selection rules leaves $82$ unique potential resonance combinations (some of which contribute both to the prograde and retrograde orbits).
This list can be further reduced by considering pro- and retrograde specific selection rules which will be discussed shortly, reducing the total amount to only $24$ resonance surfaces; $12$ for prograde and $12$ for retrograde orbits.

%\paragraph{Prograde orbits:}
%\begin{itemize}
%  \item For prograde orbits, all orbital frequencies $\omega_r$, $\omega_{\theta}$, $\omega_{\phi}$ are positive. %EVEN KIJKEN OF DE FREQUENCIES NUL KUNNEN WORDEN
%  Thus, only sets of numbers with $k < 0$ and/or $m < 0$ can lead to a resonance, so only ($n$, $k$, $m$) with $k < 0 $ and/or $m < 0$ are considered for prograde orbits.
%\end{itemize}

%\paragraph{Retrograde orbits:}
%\begin{itemize}
%  \item For retrograde orbits, $\omega_r$ and $\omega_{\theta}$ are positive and $\omega_{\phi}$ is negative. 
%  Only sets of numbers with $k < 0$ and/or $m > 0$ (note the change for $m$ compared to the prograde situation) can thus lead to a resonance, so only ($n$, $k$, $m$) with $k < 0 $ and/or $m > 0$ are considered for retrograde orbits. 
%\end{itemize}

%%%%%%%%%%%%%%%%%%%%%%%%%%%%%%%%%%%%%%%%%%
\subsubsection{Prograde orbits}
%%%%%%%%%%%%%%%%%%%%%%%%%%%%%%%%%%%%%%%%%%

Prograde orbits are defined as orbits with $0 < x < 1$. With this convention, $\omega_{\phi}$ is always positive. 

For prograde orbits, additional selection rules can be made, in addition to these mentioned in the previous section.
First, note that for prograde orbits, all orbital frequencies $\omega_r$, $\omega_{\theta}$, $\omega_{\phi}$ are positive. Then the resonance condition \eqref{eq:resonance-condition} requires $k \omega_\theta + m \omega_\phi < 0$.
Therefore, only sets of numbers with $k < 0$ and/or $m < 0$ can lead to a resonance. This condition can be succinctly written as $k+m \leq 0$. In the case of equality, $k+m = 0$, the fact that $\omega_\theta \leq \omega_\phi$ (due to Lense-Thirring nodal precession) further requires than $k>0$. 
%We will therefore only consider the cases for which
%\begin{equation}
%    k+m \leq 0.
%    \label{k+m<=0}
%\end{equation}
%In the case of equality, $k+m= 0$, we further require that $k > m$.

\begin{figure}
  \centering
  \includegraphics[width=0.46\textwidth]{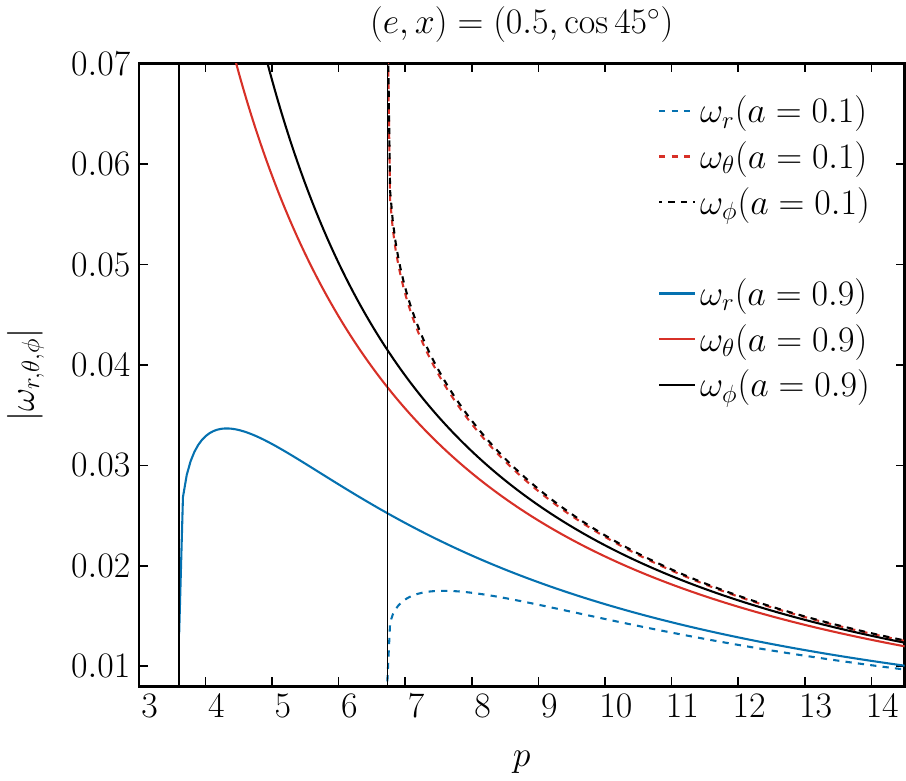}
  \caption{Prograde frequency plots showing the value of the frequencies as function of $p$ for multiple values of $a$ at fixed $e=0.5$ and inclination angle $x=\cos 50^\circ$.
  The solid lines represent the frequencies at $a = 0.9$ and the dashed lines represent the frequencies at $a = 0.1$.
  The solid and dashed vertical lines show the location of the separatrix at $a = 0.9$ and $a = 0.1$, respectively.}
  \label{fig:prograde_freqs_p_a}
\end{figure}

Secondly, for $p$ sufficiently close to the separatrix, $\omega_\theta$ and $\omega_\phi$ are significantly closer to each other than they are to $\omega_r$; see Fig.~\ref{fig:prograde_freqs_p_a}.
As a result, one can approximate the resonance condition \eqref{eq:resonance-condition} in that regime to $n \omega_r + (k+m) \omega_\theta = 0$. Additionally, since $\omega_r < \omega_\theta, \omega_\phi$, it follows that $n > |k+m|$.
%In particular, we require that 
%\begin{equation}
%    n > |k+m|.
%    \label{eq:n>k+m}
%\end{equation}

Finally, the case where $k+m = 0$ corresponds to resonance surfaces close to the separatrix. The intuition is that, near the separatrix, the radial motion ``stalls'': $\omega_r\to 0$ while the two angular frequencies remain finite and become nearly equal ($\omega_\theta\simeq\omega_\phi$); see Fig.~\ref{fig:prograde_freqs_p_a}. Writing the resonance condition $n\omega_r+k\omega_\theta+m\omega_\phi=0$ in this regime gives $n\omega_r+(k+m)\omega_\theta\simeq 0$. If $k+m\neq 0$, this can be satisfied at modest $p$ by balancing the small but nonzero $\omega_r$ against the angular term. If instead $k+m=0$, the dominant angular contribution cancels, leaving $n\omega_r\simeq 0$, which can only be met when $\omega_r$ is extremely small, i.e., for orbits very close to the separatrix. In particular, since the $n=1$ resonances require $k+m=0$, they will all appear close to the separatrix. Naturally, resonances with $k+m=0$, other than $n=1$, also exist, such as $(2,1,-1)$. However, these higher-$n$ resonances lie even closer to the separatrix than the corresponding $n=1$ resonances. They are therefore expected to be less impactful and will be ignored.

The above three conditions can be summarized as:
\begin{enumerate}[label=P\arabic*., wide]
    \item $k+m \leq 0$, where equality requires $k>0$.
    \item $n > |k+m|$, 
    \item if $k+m = 0$, then $n=1$.
\end{enumerate}
Condition P2 together with the condition that $k+m$ has to be even restricts the $n=1,2$ resonances to $k+m=0$. Together with P3, this eliminates all the $n=2$ resonances,  and restricts the $n=3,4$ resonances to $k+m = -2$.

Condition P2 also prevents any resonances with $n=0$. This may be surprising, as a priori it could be that the angular frequencies become commensurate.
To see why this does not happen, for simplicity, let us consider circular equatorial orbits for which the ratio of the fundamental frequencies takes the following simple closed form:
% \begin{align}
% \Omega_\phi &= \frac{1}{r^{3/2} + a}, \\
% \Omega_\theta^2 &= \Omega_\phi^2\left(1 - \frac{4a}{r^{3/2}} + \frac{3a^2}{r^2}\right).
% \end{align}
% Taking the ratio, we find
\begin{equation}
\frac{\omega_\theta}{\omega_\phi}
= \sqrt{1-\frac{4a}{r^{3/2}}+\frac{3a^2}{r^2}}
\le 1,
\end{equation}
with equality only as $r\to\infty$ (or for $a=0$, i.e.\ Schwarzschild, where $\omega_\theta/\omega_\phi\equiv 1$ for all radii). A resonance can occur whenever this ratio is $-m/k$.
Since $|m|\leq 2$ and the ratio is always positive, this means that the only possible resonance could be with $m/k=\{- 1/2,-1/3,-1/4,-2/3,-2/4\}$. However, given that $k+m$ has to be even this only leaves $m/k=\{-1/3,-2/4\}$, but the ratio $\omega_\theta/\omega_\phi$ only becomes $\leq 1/2$ for $r < r_{\rm separatrix}$. Hence, this does not occur for circular and equatorial orbits. 
The same argument applies for generic arguments, but is algebraically more involved.

We emphasize that the requirement P2 only holds for $p$-values sufficiently close to the separatrix. In particular, it is possible for certain resonances, such as $(2,-3,1)$ to exist at high $p$-values. For this reason, for the resonances ($n, k, m$) that are flagged by condition P2, we still look for possible resonances. However, we restrict the search for resonances to values for which $p$ is below $100$. We adopt this cutoff as a crude (and conservative) proxy for LISA relevance: resonances occurring at larger $p$ correspond to frequencies below the LISA band.\footnote{When an EMRI enters the LISA band, the semi-latus rectum is typically $p\sim 10$--$30\,M$ for a canonical $10^6\,M_\odot$ central black hole. For lighter central black holes, $\sim 10^5\,M_\odot$, the system typically does not plunge in band and instead remains at larger separations, $p\sim 50$--$100\,M$.}

Taking all these rules into account leaves us with a list of $12$ prograde resonances (see Tab.~\ref{tab:triplets}).

\begin{table}[h]
\centering
\setlength{\tabcolsep}{15pt}
\renewcommand{\arraystretch}{1.2}
\begin{tabular}{c l}
\hline
\hline
$n$ & Triplets $(k, m)$ --- Prograde orbits \\
\hline
$1$ & $(1,-1)$, $(2,-2)$ \\
$3$ & $(-4,2)$, $(-3,1)$, $(-2,0)$, $(-1,-1)$, $(0,-2)$ \\
$4$ & $(-4,2)$, $(-3,1)$, $(-2,0)$, $(-1,-1)$, $(0,-2)$ \\
\hline
\hline
\end{tabular}
\caption{Admissible triplets $(n,k,m)$ for prograde orbits.}
\label{tab:triplets}
\end{table}

\begin{figure*}[t] \centering
    \begin{subfigure}{.32\textwidth}
        \includegraphics[width=1\textwidth]{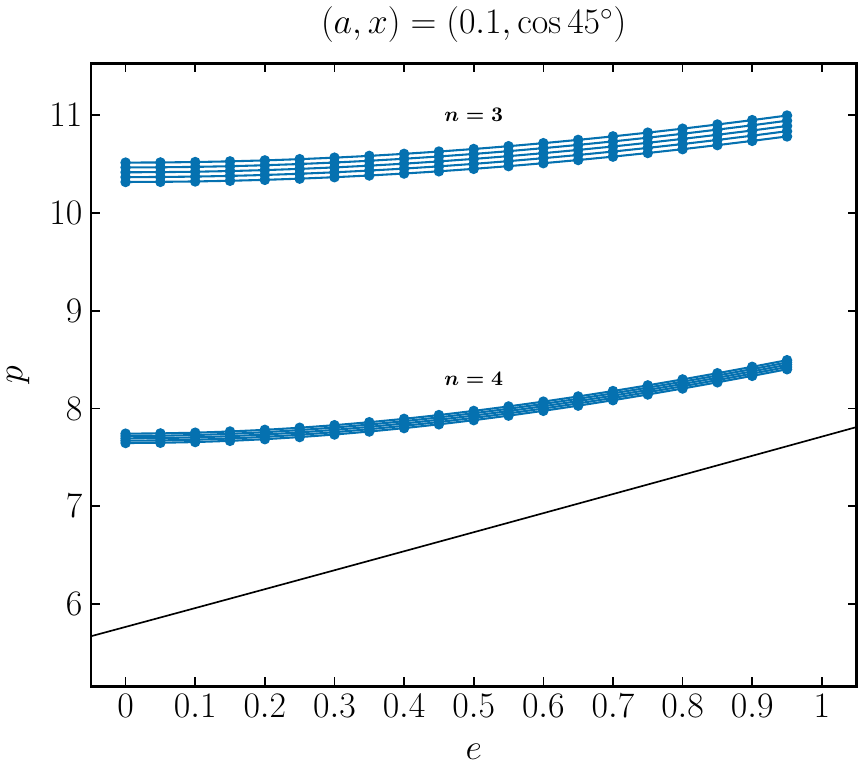}
        \centering
        \caption{}
        \label{fig:ResonanceSurface_a01_x07}
  \end{subfigure}
    \begin{subfigure}{.32\textwidth}
        \includegraphics[width=1\textwidth]{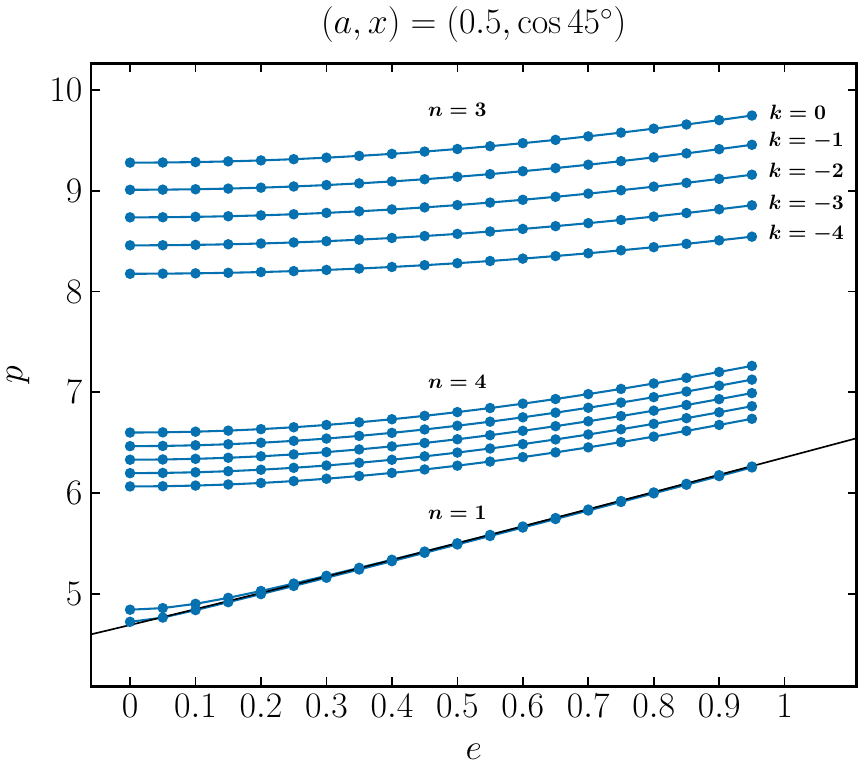}
        \centering
        \caption{}
        \label{fig:ResonanceSurface_a05_x07}
    \end{subfigure}
    \begin{subfigure}{.32\textwidth}
        \includegraphics[width=1\textwidth]{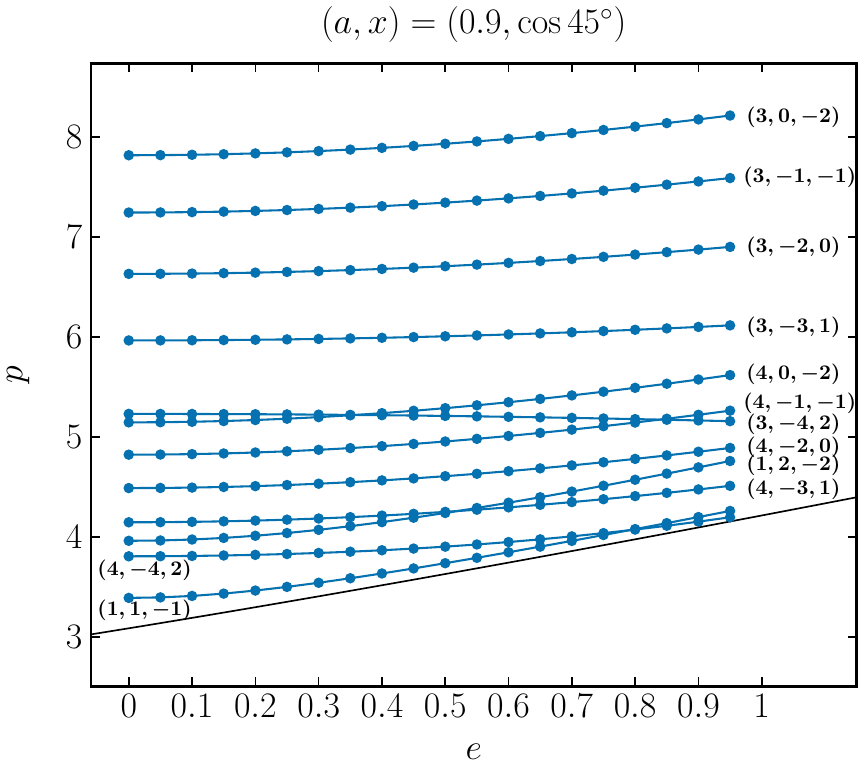}
        \centering
        \caption{}
        \label{fig:ResonanceSurface_a09_x07}
    \end{subfigure}
  \caption{Prograde resonance contours showing $p$ as a function of $e$ for fixed $a$ and $x$. Markers indicate explicitly computed resonance locations, and the curves show the corresponding interpolations.}
  \label{fig:ResonanceSurface_x07}
\end{figure*}

In Fig.~\ref{fig:ResonanceSurface_x07}, we show these 12 resonance contours at fixed inclination $x = \cos 45^\circ$, for $a=0.1$ (Fig.~\ref{fig:ResonanceSurface_a01_x07}), $a=0.5$ (Fig.~\ref{fig:ResonanceSurface_a05_x07}), and $a=0.9$ (Fig.~\ref{fig:ResonanceSurface_a09_x07}).
The two $n=1$ resonance contours are not plotted in the $a=0.1$ case Fig.~\ref{fig:ResonanceSurface_a01_x07} as they are on/slightly below the separatrix.

First, note that all resonance contours have a similar monotonic dependence on $e$, only breaking down at large spins.
In particular, for low and moderate spin ($a=0.1$ and $a=0.5$) the $n=3$ and $n=4$ create distinct resonance ``bands''. The resonances in both bands are organized in the same manner; namely the $k=0$ is the ``top'' resonance surface in a band (that is, the surface furthest away from the separatrix for a given $n$), and each subsequent surface is labeled by decreasing $k$. This is highlighted in Fig.~\ref{fig:ResonanceSurface_a05_x07}.
This feature follows from two facts: first, because $\omega_\phi > \omega_\theta$, we immediately obtain the ordering,
\begin{equation}
    \omega_{n,0,-2} > \omega_{n,-1,-1} > \omega_{n,-2,0} > \omega_{n,-3,1} > \omega_{n,-4,2},
    \label{eq:prograde:ordering}
\end{equation}
for fixed $e,x$, and small $a$.
Second, because $\omega_{nkm}$ is locally an increasing function of $p$ near $\omega_{nkm} = 0$, and the static resonance condition \eqref{eq:resonance-condition} only admits a single solution, it follows that a larger $\omega_{nkm}$ results in a smaller $p$-value; see Fig.\ref{fig:prograde_freqs_p_a}.
Similarly, the $n=3$ resonance band is higher than the corresponding $n=4$ resonance band, because both resonance surfaces in those bands satisfy $k+m=-2$, but the $n=4$ resonance have an extra $\omega_r > 0$ in $\omega_{nkm}$, resulting in a lower $p$-value. In contrast, the $n=1$ are below the $n=3,4$. The explanation for this is that they instead satisfy $k+m=0$. As mentioned before, since the angular frequencies are relatively close to each other and much larger than the radial frequency, this result in two additional factors of $\omega_\theta,\omega_\phi > \omega_r$. The weighted frequency sum $\omega_{nkm}$ is therefore larger, resulting again in a lower $p$-value.

Comparing the $a=0.1$ and $a=0.5$ cases, we see that the resonance surfaces within each band are more tightly clustered for smaller spin.
This is expected: the gap between the two angular frequencies shrinks as $a\to 0$.
In the limit $a=0$, $\omega_\theta=\omega_\phi$, so all resonances with the same $n$ and $k+m$ coincide.
This close grouping of the resonance surface within a band for small spin can pose a challenge in gravitational wave modeling with FEW, as it is trickier to correctly resolve the jumps across each resonance surface.

Finally, in Fig.~\ref{fig:ResonanceSurface_a09_x07}, we plot the resonance contours for $a=0.9$. They are consistent with the resonances modeled in previous work \cite{Gupta:2022fbe}.
Several notable differences appear in the high spin case:
First, neither the distinct bands at fixed $n$ exist anymore,  nor does the relative ordering within each band hold at high eccentricities. More dramatically, the resonance contours start to intersect at low $p$, both for different $n$, such as $(3,-4,2)$ crossing both the $(4,0,-2)$ and $(4,-1,-1)$ resonance contours, or for the same $n$, such as $(4,-3,1)$ and $(4,-4,2)$. The crossing of resonance contours poses other modeling challenges that needs addressing.
Second, while higher spin values pushes the $n=3$ and $n=4$ resonances closer to the separatrix, it seems to have the opposite effect on the $n=1$ resonances, where the $(1,1,-1)$ and $(1,2,-2)$ resonance contours are noticeably above the separatrix for $a=0.9$, very close to it at $a=0.5$, and slightly below it at $a=0.1$.
The $(1,2,-2)$ resonance contour even occurs for larger $p$ values than some of the $n = 4$ resonances. This shows that these $n = 1$ resonances can become important at high $a$.
It is expected, however, that the other $k + m = 0$ resonances are still too close to the separatrix to become important.

A general trend of Fig.~\ref{fig:ResonanceSurface_x07} shows that, as $a$ increases, all resonances happen at a lower $p$-value. 
This is because, when $a$ increases, $\omega_r$ increases while $\omega_{\theta}$ and $\omega_{\phi}$ decrease, resulting in an overall upward shift of $\omega_{nkm}$ and thus a lower resonant $p$-value.
$\omega_{\theta}$ decreases faster than $\omega_{\phi}$, so the resonance contours with a higher value of $k$ drop more. 
Note that the separatrix also decreases when $a$ increases, which is why the resonances can happen at such low $p$ for high $a$.

\subsubsection{Retrograde orbits}
%%%%%%%%%%%%%%%%%%%%%%%%%%%%%%%%%%%%%%%%%%
The above discussion only included prograde orbits. To complete the investigation of the full parameter space, the resonance contours for retrograde orbits will be discussed in this subsection.
The retrograde orbits are classified as orbits with $-1 < x < 0$. With this convention, $\omega_{\phi}$ becomes negative. 

\begin{figure}[h]
  \centering
  \includegraphics[width=0.47\textwidth]{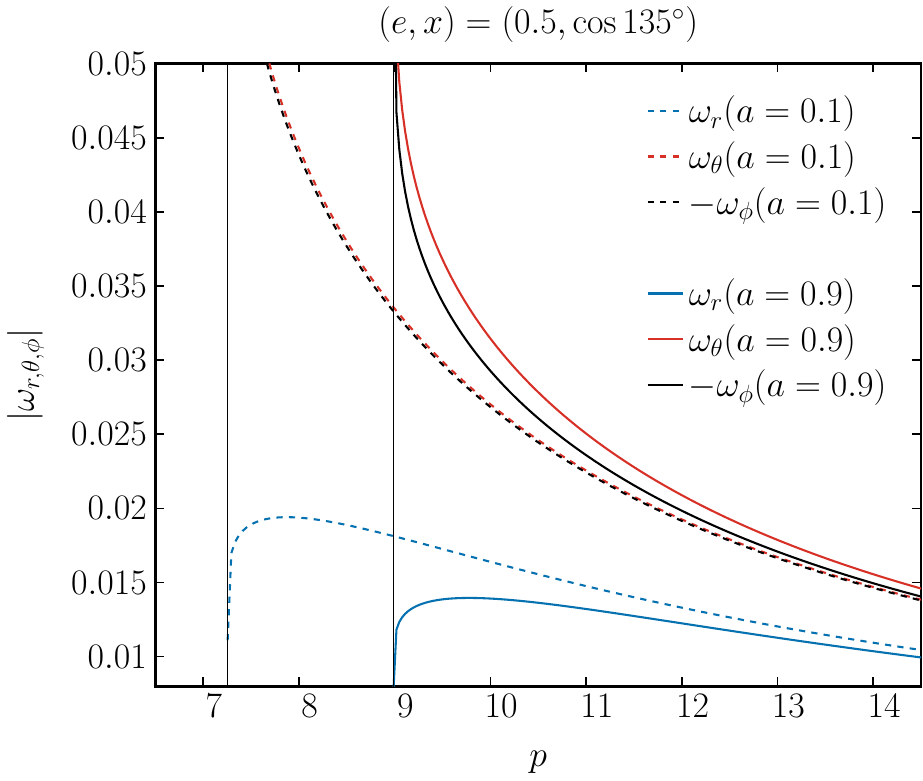}
  \caption{Retrograde frequency plots showing the value of the frequencies as function of $p$ for multiple values of $a$. 
  The solid lines represent the frequencies at $a = 0.9$ and the dashed lines represent the frequencies at $a = 0.1$. 
  The solid and dashed vertical lines show the location of the separatrix for $a = 0.9$ and $a = 0.1$, respectively. }
  \label{fig:retrograde_freqs_p_a}
\end{figure}

There are two main differences between prograde and retrograde orbits. First, as mentioned earlier, $\omega_\phi$ is negative for retrograde orbits. 
Second, looking at Fig.~\ref{fig:retrograde_freqs_p_a}, it can be seen that $|\omega_\theta|>|\omega_\phi|$ for retrograde orbits, while for prograde orbits, $|\omega_\theta|<|\omega_\phi|$.

These differences between prograde and retrograde orbits result in key difference in their corresponding selection rules.
First, since $\omega_r >0$, we still require that $k \omega_\theta +m \omega_\phi < 0$. Since $\omega_\phi < 0$ for retrograde orbits, this enforces $k < 0$ and/or $m > 0$, or more succinctly, $k-m \leq 0$.
In the case of equality, the switch in the relative magnitudes between $\omega_\theta$ and $\omega_\phi$ then requires $k < 0$.
Second, as can be seen in Fig.~\ref{fig:retrograde_freqs_p_a}, $\omega_\theta$ and $-\omega_\phi$ are closer to each other than they are to $\omega_r$. So, the resonance condition \eqref{eq:resonance-condition} can be approximated by $n \omega_r + (k-m) \omega_{\theta} = 0$, from which we derive the condition $n > |k-m|$. 
Finally, similarly as for the prograde orbits, $k-m = 0$ corresponds to resonance contours close to the separatrix and of those the $n=1$ resonances are the furthest away from the separatrix, which is why we only consider these.

The above three conditions can be summarized as:
\begin{enumerate}[label=R\arabic*., wide]
    \item $k-m \leq 0$, where equality requires $k<0$.
    \item $n > |k-m|$,
    \item if $k-m = 0$, then $n=1$. 
\end{enumerate}
As for prograde orbits, condition R2 restricts the $n=1,2$ resonances to $k-m=0$. Together with R3, this eliminates all the $n=2$ resonances, and restricts the $n=3,4$ resonances to $k-m = -2$.

These additional selection rules result in the $12$ retrograde resonances listed in Tab.~\ref{tab:triplets-retrograde}. This list of retrograde orbits match those for prograde orbits, by switching $k \to -k$ for $n=1$ and $m \to -m$ for $n=3,4$.

\begin{table}[h]
\centering
\setlength{\tabcolsep}{15pt}
\renewcommand{\arraystretch}{1.2}
\begin{tabular}{c l}
\hline
\hline
$n$ & Triplets $(k, m)$ --- Retrograde orbits \\
\hline
$1$ & $(-1,-1)$, $(-2,-2)$ \\
$3$ & $(-4,-2)$, $(-3,-1)$, $(-2,0)$, $(-1,1)$, $(0,2)$ \\
$4$ & $(-4,-2)$, $(-3,-1)$, $(-2,0)$, $(-1,1)$, $(0,2)$ \\
\hline
\hline
\end{tabular}
\caption{Admissible triplets $(n,k,m)$ for retrograde orbits.}
\label{tab:triplets-retrograde}
\end{table}

\begin{figure*}[t] \centering
    \begin{subfigure}{.32\textwidth}
        \includegraphics[width=1\textwidth]{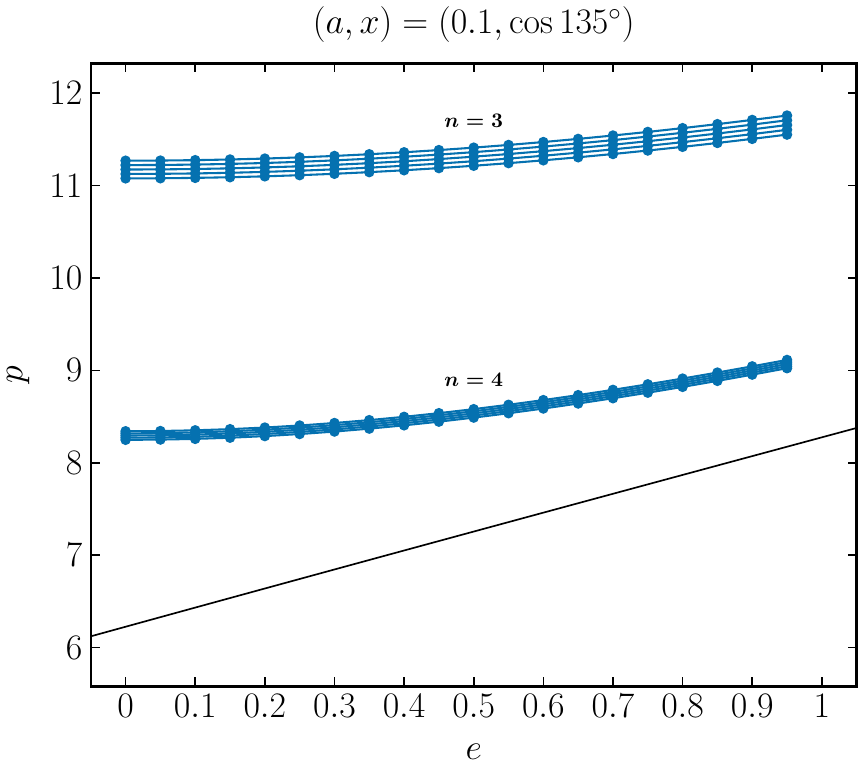}
        \centering
        \caption{}
        \label{fig:ResonanceSurface_a01_x-07}
  \end{subfigure}
    \begin{subfigure}{.32\textwidth}
        \includegraphics[width=1\textwidth]{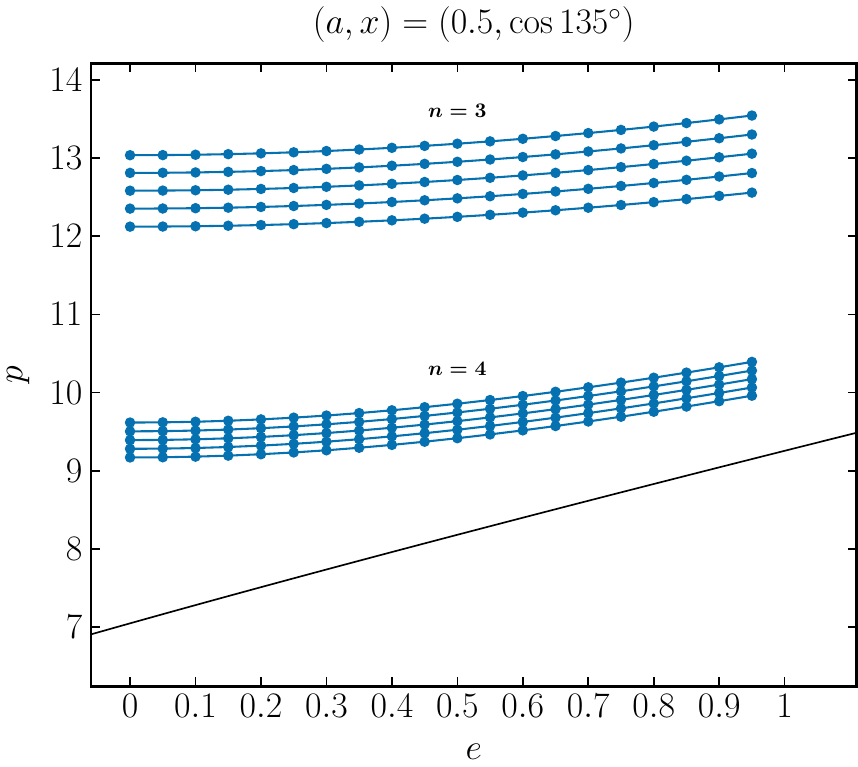}
        \centering
        \caption{}
        \label{fig:ResonanceSurface_a05_x-07}
    \end{subfigure}
    \begin{subfigure}{.32\textwidth}
        \includegraphics[width=1\textwidth]{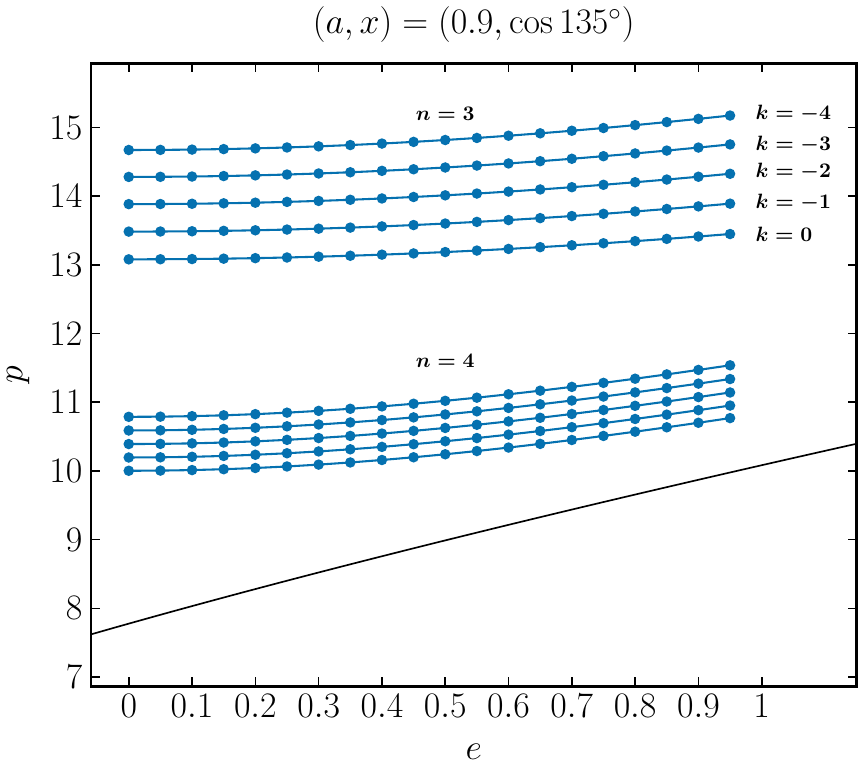}
        \centering
        \caption{}
        \label{fig:ResonanceSurface_a09_x-07}
    \end{subfigure}
  \caption{Retrograde resonance contours showing the $p$ value as a function of $e$ for fixed $a$ and $x$.}
  \label{fig:ResonanceSurface_x-07}
\end{figure*}

In Fig.~\ref{fig:ResonanceSurface_x-07}, we show all of the resonance contours, at fixed inclination $x = \cos 135^\circ$, for $a=0.1$ (Fig.~\ref{fig:ResonanceSurface_a01_x-07}), $a=0.5$ (Fig.~\ref{fig:ResonanceSurface_a05_x-07}), and $a=0.9$ (Fig.~\ref{fig:ResonanceSurface_a09_x-07}). Unlike for the prograde case, the $n=1$ resonances are all slightly above or below the separatrix for these cases, and are therefore not shown.
%In particular, the $a=0.5$ case agrees with previous work~\cite{guptaModelingTransientResonances2022}.

Comparing Fig.~\ref{fig:ResonanceSurface_x07} with Fig.~\ref{fig:ResonanceSurface_x-07}, we highlight several similarities and distinctions between the prograde and retrograde cases.

Among the similarities, the resonance contours display broadly the same monotonic dependence on the eccentricity $e$. This is particularly true for small values of $a$. For small $a$, we also observe the same band structure at different $n$, with the resonance contours within each band moving closer together as the spin decreases. As in the prograde case, this is because in the Schwarzschild limit, $a=0$, we have $\omega_\theta = -\omega_\phi$, so that resonances with the same $k-m$ overlap.

Among the distinctions, we first note that the ordering within each band is reversed relative to the prograde case. This follows from the fact that in the retrograde case $|\omega_\theta| > |\omega_\phi|$, so that the prograde ordering in \eqref{eq:prograde:ordering} is essentially reversed (together with sending $m \to -m$).

A second difference is that the retrograde resonances occur at larger $p$ values than the prograde ones. This is due to frame-dragging now acting against the orbit, which pushes the separatrix and all resonance contours to higher $p$. This shift can also be seen in Fig.~\ref{fig:retrograde_freqs_p_a}, where $\omega_r$, $\omega_\theta$ and $-\omega_\phi$ are shown as a function of $p$ for $e = 0.5$ and $x = \cos 135^\circ$.
%It can be seen in figure \ref{fig:retrograde_freqs_p_a} that the functional dependence of the frequencies on $p$ is virtually the same as for the prograde orbits in figure \ref{fig:prograde_freqs_p_a}, but shifted. 

Perhaps the most distinctive feature is that the $n=3$ and $n=4$ band structure persists at high spin in the retrograde case. This persistence arises for two reasons: (1) as $a$ increases, the resonance contours shift to higher $p$ (rather than to lower $p$, as for prograde orbits), and (2) the contours for a given $n$ spread apart less with increasing $a$. The shift to higher $p$ with increasing $a$ follows from the fact that, for retrograde orbits, $\omega_r$ decreases while $\omega_\theta$ and $-\omega_\phi$ increase as $a$ increases. This produces an overall decrease in $\omega_{nkm}$ and hence a higher $p$ value for the resonance.

Why the contours for a given $n$ remain close together despite the high $a$ values is less clear. One possible explanation is that, because the retrograde resonance contours lie farther from the central BH, the gravitational influence is weaker, so the contours behave more similarly and stay grouped together.

%As most features of the frequencies are the same for both prograde and retrograde orbits, except for $\omega_{\phi}$ becoming negative,
%the $n = 3$ and $n = 4$ resonances appear for the same $k$ as for prograde orbits, but accompanied by the negative of the $m$ of the combination for prograde orbits. 
%The $n = 1$ resonances are different, however, and occur at $(1, -k, -m)$ compared to their prograde counterparts. This is because $|\omega_{\phi}| < |\omega_{\theta}|$ for retrograde orbits and $|\omega_{\phi}| > |\omega_{\theta}|$ for prograde orbits. 
%The flipped sign in the inequality results in resonances occurring for $-k$ and $-m$. 
%Looking at figure \ref{fig:retrograde_freqs_p_a}, it can be seen that $\omega_{\theta}$ is larger than $-\omega_{\phi}$ for retrograde orbits. %WHY IS THIS??%
%The result of this is that the order of the resonances in a band is flipped in comparison to the prograde resonances contours. 
%The resonances ($n$, 0, 2) appear now at the lowest $p$ and the resonances with ($n$, $-4$, $-2$) appear at the highest $p$.\\

%Note that, for retrograde orbits, increasing the inclination angle $I$ is equivalent to moving towards the equatorial plane, so for both prograde and retrograde orbits, 
%the resonance contours increase with increasing $I$. 

%%%%%%%%%%%%%%%%%%%%%%%%%%%%%%%%%%%%%%%%%%
\subsubsection{Dependence on the inclination angle}
%%%%%%%%%%%%%%%%%%%%%%%%%%%%%%%%%%%%%%%%%%
So far we have discussed only how the resonance contours depend on $a$ and $e$. Their position, however, also depends on the inclination, parametrized by $x = \cos I$. At fixed $a$ and $e$, the contours at different inclinations remain qualitatively very similar to those in Fig.~\ref{fig:ResonanceSurface_x07} (prograde) and Fig.~\ref{fig:ResonanceSurface_x-07} (retrograde); the only significant change is that they shift to higher $p$ as $x$ decreases.
%driven by the fact that $\omega_\theta$ and $|\omega_\phi|$ are increasing functions of $x$.
%Resulting in $\omega_{nkm}$ shifting downwards and the resonance contours shifting upwards.
Equivalently, the contours move to higher $p$ as the orbit tilts away from the equatorial plane for prograde orbits, and to lower $p$ as it tilts away for retrograde orbits. Both trends share a common origin in frame-dragging, whose coupling to the orbit is strongest in the equatorial plane and weakens towards polar orbits. For prograde orbits, frame-dragging assists the motion of $\mu$, pulling the contours to lower $p$; this assistance is maximal at $x=1$ and diminishes as the orbit inclines, so the contours move back up. For retrograde orbits, frame-dragging instead opposes the motion, pushing the contours to higher $p$; this opposition is maximal at $x=-1$ and likewise weakens with inclination, so the contours move back down. Across the full range, the net result is the monotonic shift to higher $p$ with decreasing $x$ noted above.

This inclination dependence strengthens with $a$, as expected: in the Schwarzschild limit the spacetime is spherically symmetric, so the frequencies, and hence the resonance contours, must be independent of $I$. 
%The entire effect is therefore controlled by the spin.

%%%%%%%%%%%%%%%%%%%%%%%%%%%%%%%%%%%%%%%%%%
\subsubsection{High $n$ resonance surfaces}
%%%%%%%%%%%%%%%%%%%%%%%%%%%%%%%%%%%%%%%%%%

\begin{figure*}
  \centering
  \begin{subfigure}{.49\textwidth}
      \includegraphics[width=0.95\textwidth]{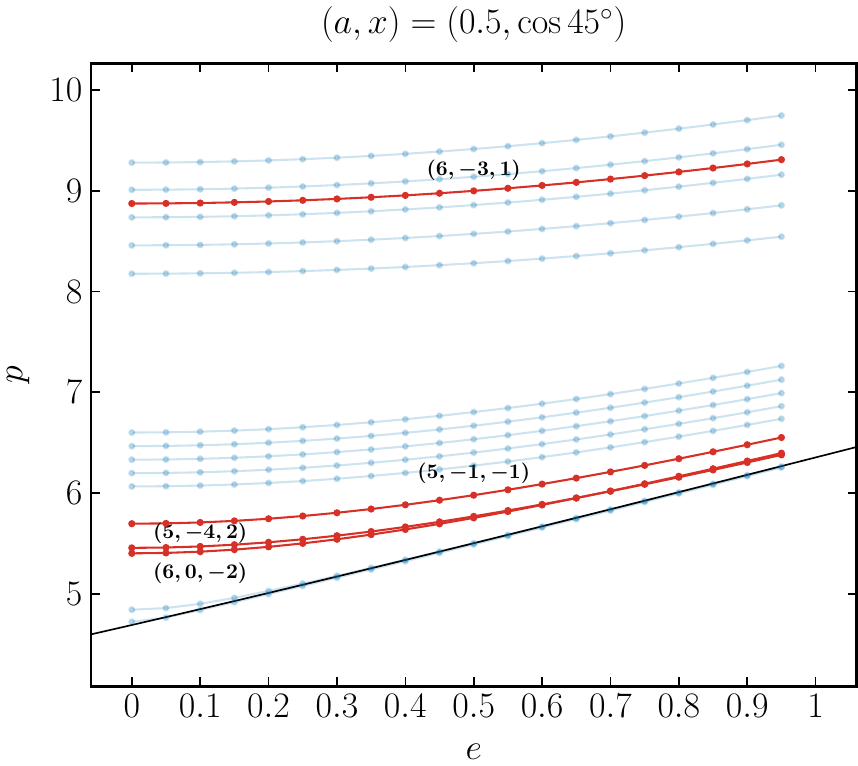}
      \centering
      \caption{}
      \label{fig:extracontours_prograde}
  \end{subfigure}
  \begin{subfigure}{.49\textwidth}
      \includegraphics[width=0.95\textwidth]{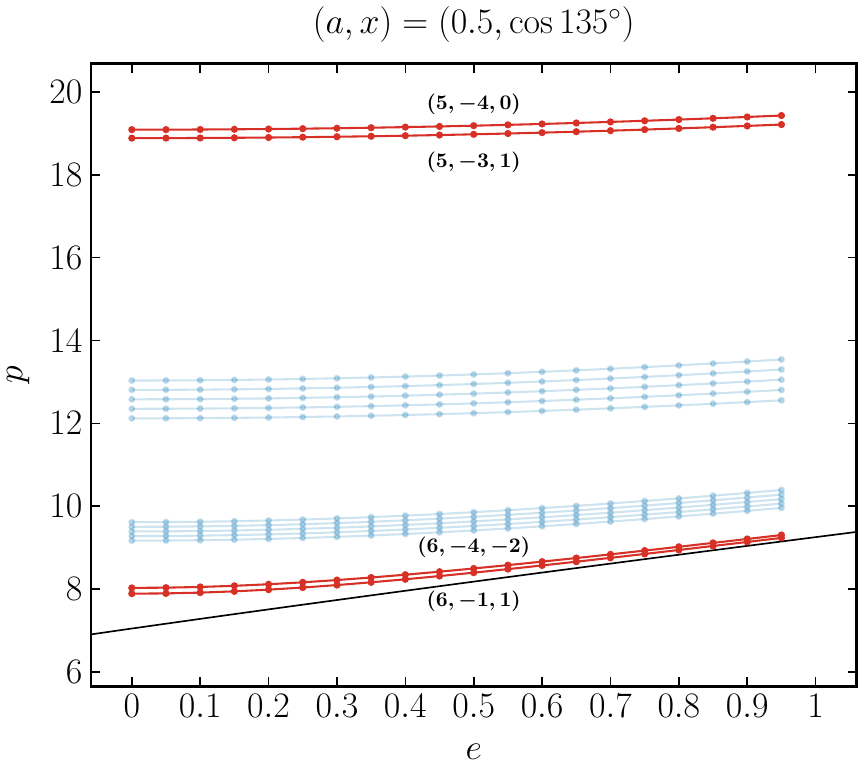}
      \centering
      \caption{}
      \label{fig:extracontours_retrograde}
      \end{subfigure}
  \caption{Resonance contours showing the resonant $p$ value as function of $e$. The $n = 5$ and $n = 6$ resonances are highlighted. 
   Prograde resonances at $a = 0.5$ and $I = 50^\circ$ are shown in (a) and retrograde resonances at $a = 0.5$ and $I = 130^\circ$ are shown in (b).}
\label{fig:extracontours}
\end{figure*}

Much of our analysis so far has been restricted to $n\leq4$. To confirm that the features and trends we identified are not peculiar to these lower resonances, we also computed contours for two $n=5$ and two $n=6$ resonances. In Fig.~\ref{fig:extracontours} we show these alongside the $n \leq 4$ contours (faded), for prograde (\ref{fig:extracontours_prograde}) and retrograde (\ref{fig:extracontours_retrograde}) orbits. In both panels the spin is fixed to $a = 0.5$.

Consider first the prograde orbits. The $(5,-1,-1)$, $(5,-4,2)$, and $(6,0,-2)$ resonances exhibit the same behavior as their $n=3$ and $n=4$ counterparts. First, they all lie between the $n=4$ contours and the separatrix: since these resonances share the same $k+m=-2$, a higher $n$ yields a higher $\omega_{nkm}$ and hence a smaller $p$. Second, as for the lower resonances, $(5,-1,-1)$ sits at a larger $p$ than $(5,-4,2)$; this again follows because for prograde orbits $\omega_\phi > \omega_\theta$, so at fixed $k+m$ a larger $m$ (equivalently, smaller $k$) gives a larger $\omega_{nkm}$.

This leaves the $(6,-3,-1)$ resonance, which has $k+m=-4$ and so does not belong to the same ``family''. Its location is readily understood from the observation that $(n,k+m) = (6,4) = 2 \times (3,2)$, so this resonance should appear near the $n=3$ contours. More precisely, $(6,-3,-1) = 2 \times (3,-1.5,-0.5)$. Formally, the resonant $p$-value of a resonance $(n,k,m)$ coincides with that of $(\frac{n}{\lambda},\frac{k}{\lambda},\frac{m}{\lambda})$ for any $\lambda \in \mathbb{N}$, so $(6,-3,-1)$ shares the $p$-value of $(3,-1.5,-0.5)$. This latter ``resonance'' has $n=3$ and $k+m=-2$, placing it on the same footing as the other resonances with that $(n,k+m)$ pair. Using once more that $\omega_\phi > \omega_\theta$, we conclude that $(6,-3,-1)$ must lie between $(3,-1,-1)$ and $(3,-2,0)$, in agreement with Fig.~\ref{fig:extracontours_prograde}.

Although hard to discern in Fig.~\ref{fig:extracontours_prograde}, the $(5,-4,-2)$ and $(6,0,-2)$ contours intersect. Since we have modeled only a handful of $n=5$ and $n=6$ contours, including the full set would reveal further intersections, and we expect more to appear at higher $a$.

The retrograde high-$n$ surfaces, Fig.~\ref{fig:extracontours_retrograde}, display the same features found for the $n\leq4$ resonances: for a given $k-m$, the $n=6$ resonance lies at a smaller $p$ than the $n=5$ resonance. The $n=5$ resonances with $k-m=-4$, however, appear at a markedly higher $p$ than all the others. As in the prograde case, this is because $(5,-4,0)$ shares its $p$-value with $(2.5,-2,0)$: this surface lies above $(3,-2,0)$ since $\omega_r > 0$ gives it a smaller $\omega_{nkm}$, and hence a larger $p$. The reason $(5,-4,0) \sim (2.5,-2,0)$ sits so much higher than $(3,-2,0)$ is that its $n$ is much closer to its $|k-m|$: the smaller the gap between $n$ and $|k-m|$, the higher the resonant $p$-value. In the limit $n \to |k-m|$ (or $n \to k+m$ for prograde orbits), $p$ diverges, as all three orbital frequencies must then approach their common Keplerian value; see Eq.~\ref{eq:Keplerian}.

More generally, we can estimate where a resonance contour falls by computing the ratio
\begin{equation}
  \rho = \frac{|k \pm m|}{|n|},
  \label{eq:rationandkpm}
\end{equation}
with the $+$ for prograde resonances and the $-$ for retrograde ones. The closer $\rho$ is to $1$, the higher the $p$-value of the contour; a $\rho$ near $0$ indicates a contour close to the separatrix. This measure relies on the approximation $\omega_\theta \approx |\omega_\phi|$, and so Eq.~\ref{eq:rationandkpm} breaks down at high $a$.

%%%%%%%%%%%%%%%%%%%%%%%%%%%%%%%%%%%%%%%%%%
%%%%%%%%%%%%%%%%%%%%%%%%%%%%%%%%%%%%%%%%%%
\section{Resonance durations}
\label{sec:duration}
%%%%%%%%%%%%%%%%%%%%%%%%%%%%%%%%%%%%%%%%%%
%%%%%%%%%%%%%%%%%%%%%%%%%%%%%%%%%%%%%%%%%%
In this section, we compute the resonance duration in Eq.~\eqref{eq:ResDuration} using the FastEMRIWaveform (FEW) package. We first compare a post-Newtonian approximation to the self-force results for $T^{\rm{res}}_{nkm} $, before investigating general trends.

We first compute $\dot{\omega}_{r,\theta,\phi}$ with  FEW, where as before the overdot denotes a time-derivative with respect to Boyer-Lindquist time. FEW readily returns the derivatives of the orbital parameters $\dot{p}, \dot{e}, \dot{x}$ for a given quadruplet $(a,p,e,x)$ (the spin evolution $\dot{a}$ is a second-order effect and is neglected), and there is a one-to-one map $(\dot{p},\dot{e},\dot{x}) \leftrightarrow (\dot{\omega}_r, \dot{\omega}_\theta, \dot{\omega}_\phi)$. FEW offers two modules for computing $(\dot{p},\dot{e},\dot{x})$: ``PN5'' which evolves the trajectory using a 5th-order post-Newtonian approximation, and ``Kerr'' which uses the exact Kerr expressions (to linear order in the self-force). Since the Kerr module currently supports only equatorial trajectories, all Kerr/PN5 comparisons below are restricted to $x=\pm 1$.

\begin{figure*}
  \centering
  \begin{subfigure}{.49\textwidth}
      \includegraphics[width=0.9\textwidth]{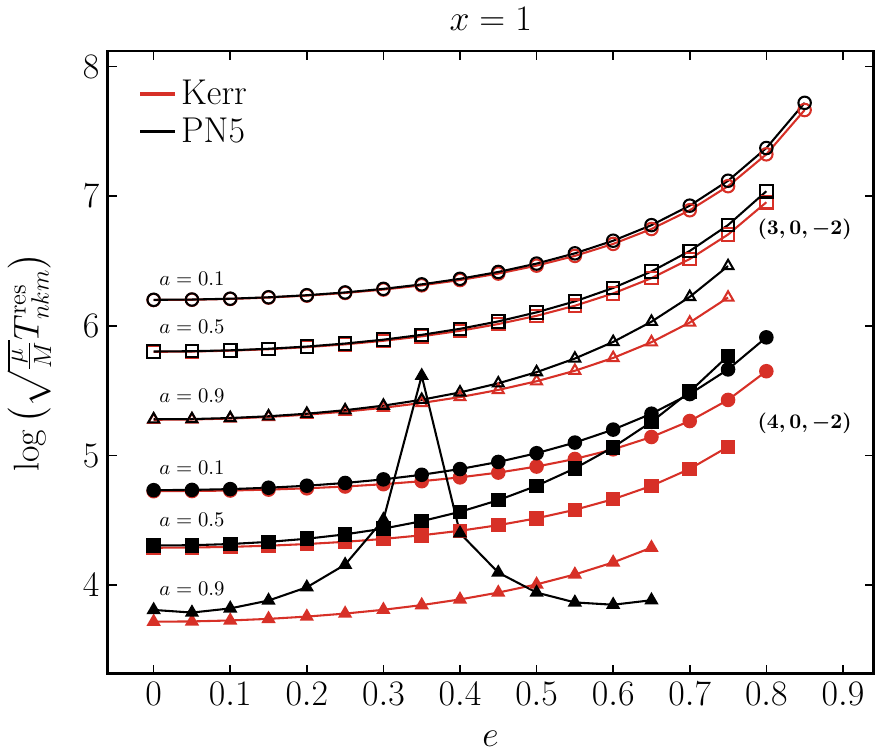}
      \centering
      \caption{}
      \label{fig:Duration2D_All_x1}
  \end{subfigure}
  \begin{subfigure}{.49\textwidth}
      \includegraphics[width=0.9\textwidth]{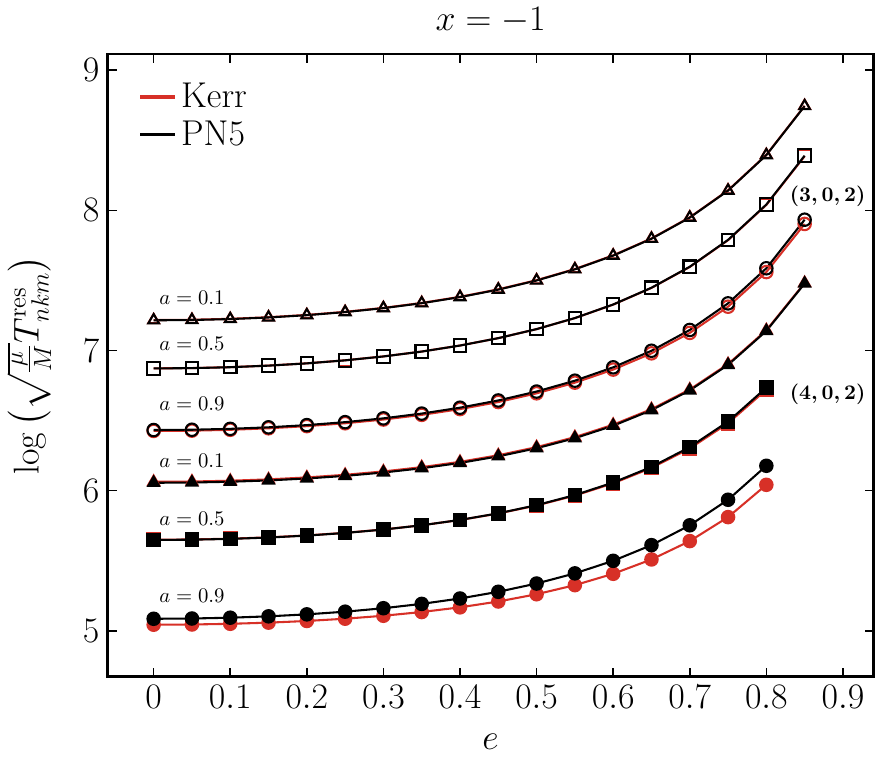}
      \centering
      \caption{}
      \label{fig:Duration2D_All_x-1}
      \end{subfigure}
  \caption{Comparison of the exact Kerr( red) and PN5 (black) computations of the resonance duration $T^{\rm res}_{nkm}$ (rescaled by the square root of the mass-ratio), for equatorial prograde $x=1$ (a) and retrograde $x=-1$ (b) orbits. To avoid clutter, we show only the $(3,0,\mp 2)$ and $(4,0,\mp 2)$ resonance surfaces; the others are qualitatively similar.}
  \label{fig:Duration2D_All}
\end{figure*}

Figure~\ref{fig:Duration2D_All} compares the resonance durations, Eq.~\ref{eq:ResDuration}, from the exact Kerr and PN5 calculations for the $(3,0,-2)$ and $(4,0,-2)$ resonances at spins $a=0.1,\,0.5,$ and $0.9$. Owing to the Kerr restriction noted above, we consider only equatorial prograde (Fig.~\ref{fig:Duration2D_All_x1}) and retrograde (Fig.~\ref{fig:Duration2D_All_x-1}) orbits.

For both prograde and retrograde orbits, PN5 approximates the Kerr result well at low eccentricity, low spin, and for resonances far from the separatrix. The largest disagreement therefore occurs for the $n=4$ resonances at $a=0.5$ and $a=0.9$, peaking at high eccentricity. The one striking exception is the $(4,0,-2)$ resonance at $a=0.9$ for prograde orbits, where the PN5 duration spikes near $e \approx 0.35$; we verified that the other $n=4$ resonances show a similar peak. We could not pin down its precise cause, but attribute it to the breakdown of the PN5 approximation at this combination of high spin and proximity to the separatrix.

PN5 and Kerr agree noticeably better for retrograde orbits; in particular, the anomalous spike seen for $(4,0,-2)$ at $a=0.9$ is absent. This is because these resonances occur at larger $p$ values, i.e., deeper in the weak-field regime where the PN expansion is more accurate. The larger $p$ values also account for the longer durations: the corresponding retrograde resonances last roughly an order of magnitude longer than their prograde counterparts.

Figure~\ref{fig:Duration2D_All} not only enables a direct comparison between PN5 and exact Kerr, but also reveals several general trends. First, the resonance duration depends strongly on spin: low-spin cases typically last about an order of magnitude longer than high-spin cases. Second, the duration increases with eccentricity, with larger $e$ leading to longer resonances. Third, the $(3,0,-2)$ resonance lasts roughly an order of magnitude longer than the $(4,0,-2)$ resonance. More generally, $n=3$ resonances occur at larger $p$ values and have durations about an order of magnitude larger than $n=4$ resonances.

These trends are a key driver in determining the relative importance of different resonances, since longer durations generally correspond to stronger resonances. Although the final jump size is mediated by the resonance strength, we will see that smaller spins and higher eccentricities typically lead to larger jumps.

\begin{figure*}[h]
    \centering
    \begin{subfigure}[b]{0.99\textwidth}
        \centering
        \includegraphics[width=0.9\textwidth]{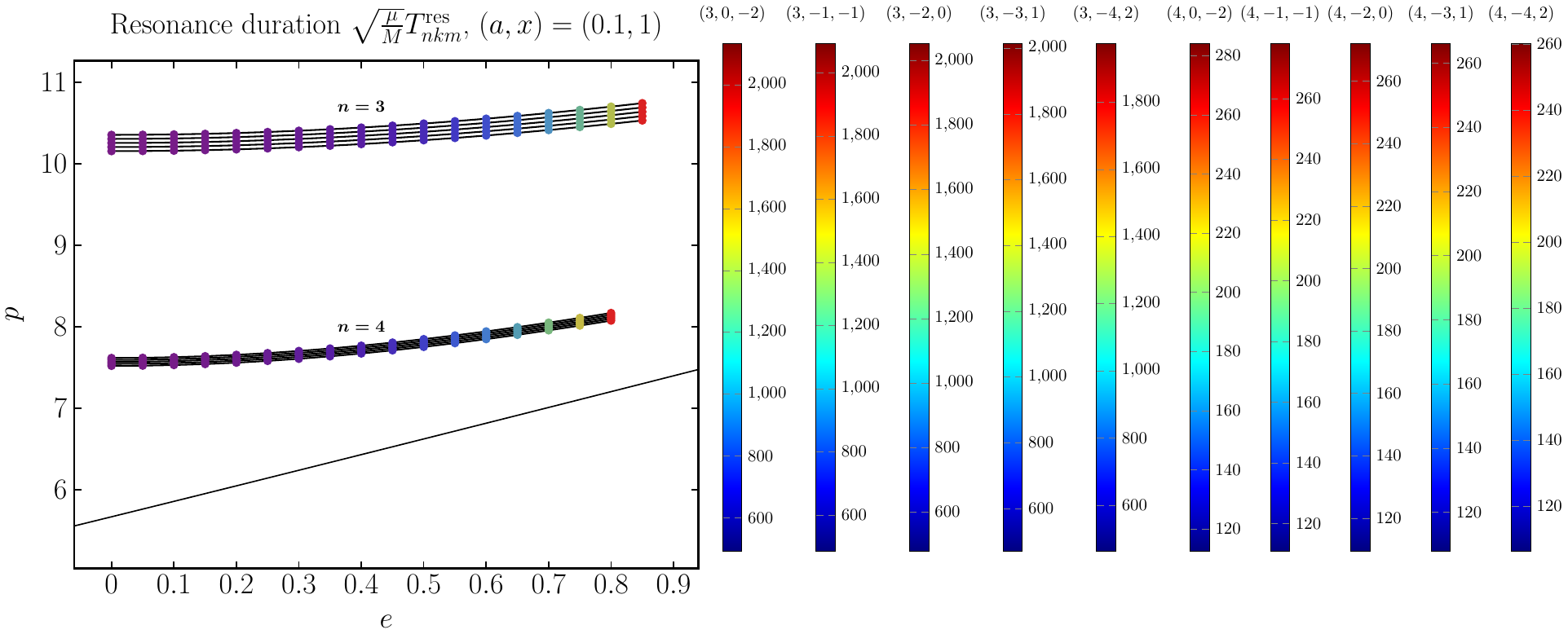}
        \caption{}
        \label{subfig:plot1}
    \end{subfigure}\hfill
    \vspace{0.25cm} % Vertical gap between rows
    \begin{subfigure}[b]{0.99\textwidth}
        \centering
        \includegraphics[width=0.9\textwidth]{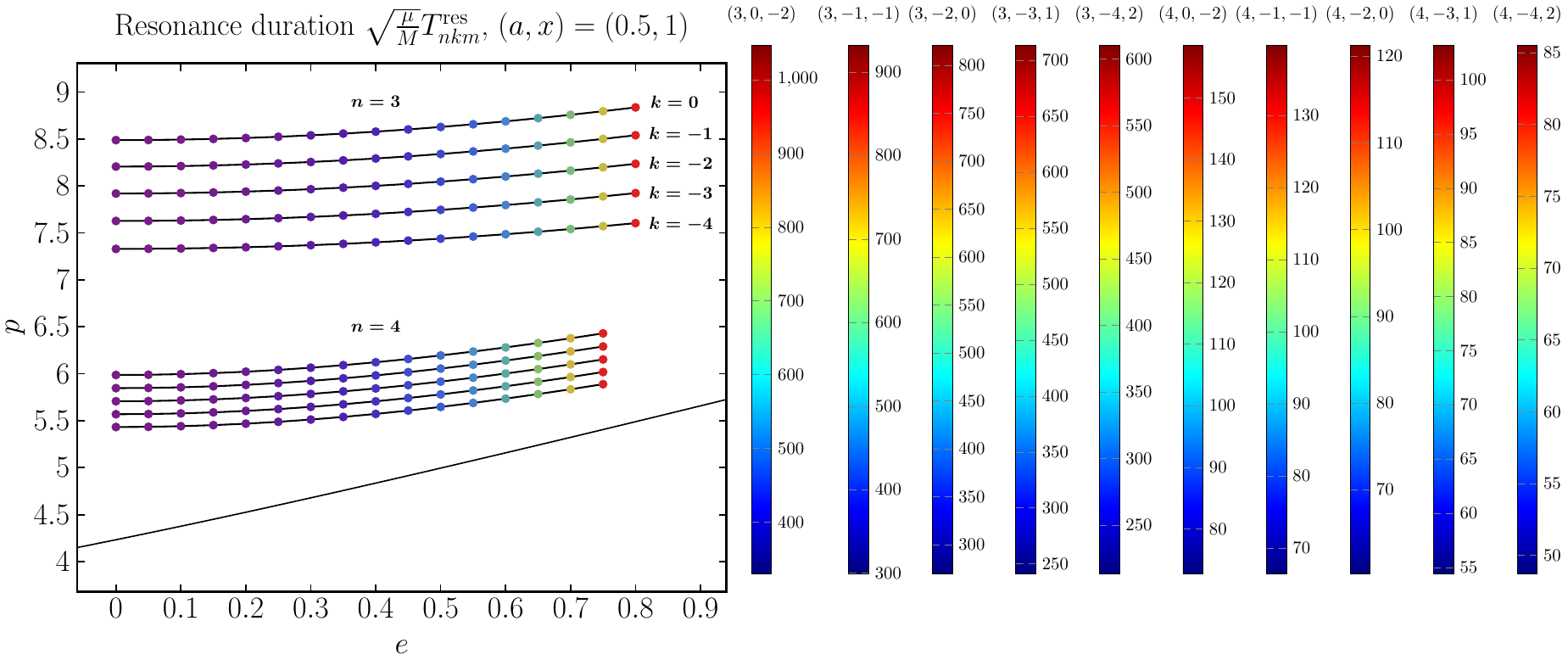}
        \caption{}
        \label{subfig:plot2}
    \end{subfigure}\hfill
    \vspace{0.25cm} % Vertical gap between rows
    \begin{subfigure}[b]{0.99\textwidth}
        \centering
        \includegraphics[width=0.9\textwidth]{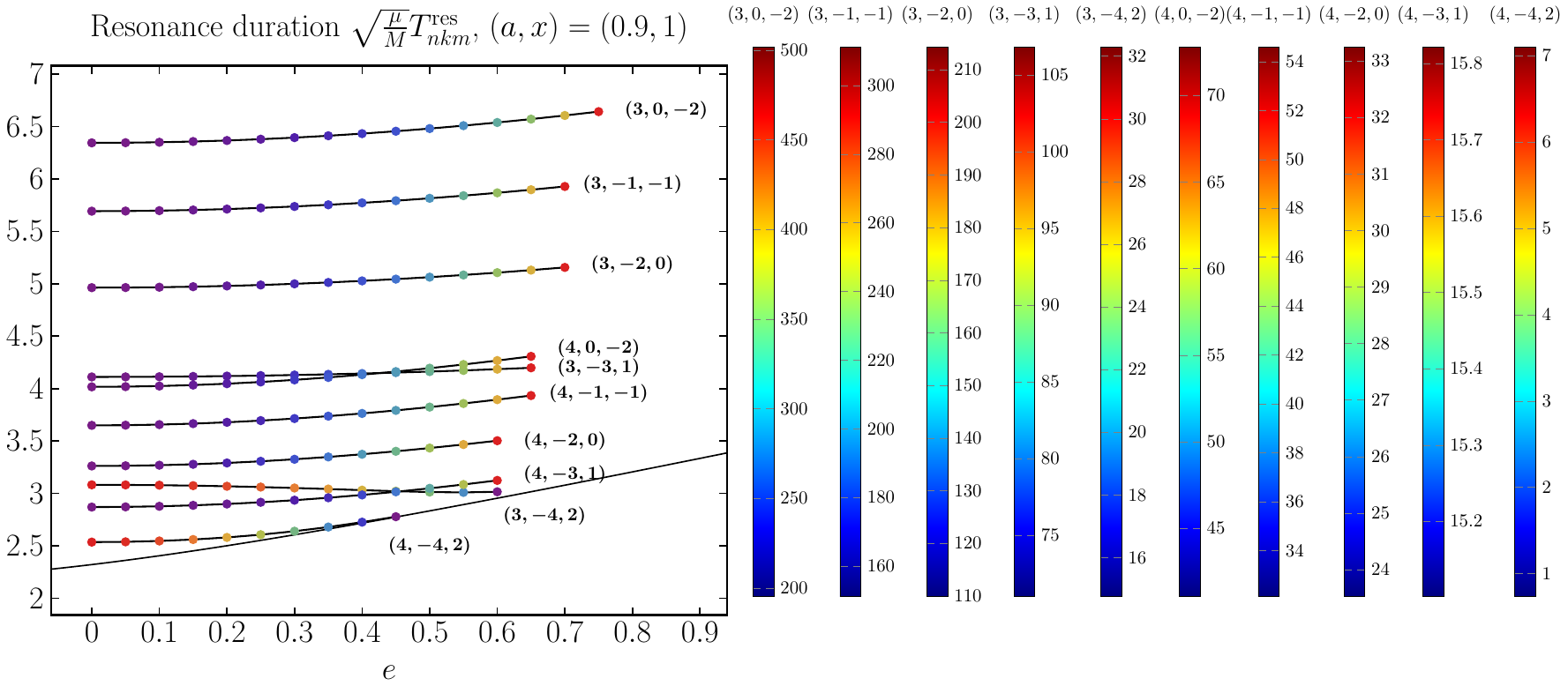}
        \caption{}
        \label{subfig:plot3}
    \end{subfigure}
    \caption{Heat map of the resonance surface duration scaled by the mass ratio: $\sqrt{\tfrac{\mu}{M}}T^\text{res}_{nkm}$ \eqref{eq:ResDuration} for equatorial prograde orbits, $x=1$, for spin $a=0.1$ (a), $a=0.5$ (b) and $a=0.9$ (c) for all the resonances in Tab.~\ref{tab:triplets}, with the exception of the two $n=1$ resonances, as they are very close to the separatrix.}
    \label{fig:DurationColorMap}
\end{figure*}

Figure~\ref{fig:DurationColorMap} shows the resonance duration \eqref{eq:ResDuration} for all resonances as a heat map, using the exact Kerr data for equatorial prograde orbits ($x=1$). Two features stand out. First, the duration generally increases with eccentricity, except where the resonance contour approaches the separatrix. Second, the duration decreases with increasing $n$ and decreasing $k$---again a consequence of the contour's proximity to the separatrix. Physically, both trends are consistent with the expectation that the resonance duration vanishes on the separatrix. This is exemplified by the $(3,-4,2)$ and $(4,-4,2)$ resonances in the high spin case: these are the only modes whose duration decreases with eccentricity because they approach the separatrix as eccentricity increases. Likewise, the $(4,-3,1)$ duration in the high spin case still increases with $e$, but over a notably small range. Together, these observations point to a competition between two effects: the intrinsic growth of the resonance duration with eccentricity, and its suppression as the resonant $p$-value nears the separatrix.
For completeness, Fig.~\ref{fig:OmegaDotColorMap} in App.~\ref{app:omega-dot} shows the individual contributions to $T^{\rm res}_{nkm}$ for an inclined orbit.

With regards to the dependence of $T^{\rm res}_{nkm}$ to inclination, we find that $T^{\rm res}_{nkm}$ generally decreases with increasing $x$, for both pro- and retro-grade orbits. In terms of the inclination angle, this means that, as one approaches an equatorial orbit, $T^{\rm res}_{nkm}$ decreases for prograde orbits, while it increases for retrograde orbits.

The dependence of the quantity on the inclination angle is strongly modulated by the spin parameter, and to a lesser extent by the eccentricity. For small values of the spin parameter $a$, the variation across inclination angles is relatively modest. For instance, for $a = 0.1$, the difference between the smallest and largest values is typically on the order of $\sim 10\text{-}20\%$ at most. In contrast, for larger values of $a$ (here assessed up to $a=0.9$), the ratio between the smallest and largest values increases substantially, reaching factors of approximately $2$-$3$.

%%%%%%%%%%%%%%%%%%%%%%%%%%%%%%%%%%%%%%%%%%
%%%%%%%%%%%%%%%%%%%%%%%%%%%%%%%%%%%%%%%%%%
\section{Jump amplitudes} \label{sec:jumpamps}
%%%%%%%%%%%%%%%%%%%%%%%%%%%%%%%%%%%%%%%%%%
%%%%%%%%%%%%%%%%%%%%%%%%%%%%%%%%%%%%%%%%%%
The last ingredient is the jump amplitude. The net impact of a resonance scales as ``(jump amplitude)$\times$(resonance duration)$\times e^{i \, \text{phase}}$''; since the phase is highly sensitive to the initial conditions, we focus on the phase-independent quantity ``(jump amplitude)$\times$(resonance duration)'' to compare the relative importance of different resonances. The underlying jump-amplitude and duration data are publicly available  \cite{GitHubRepo}.

Because the tidal forcing is time-translation invariant in our stationary-perturber model, there is no jump in $E$. We verified that our computed $\Delta E$ is consistent with numerical noise, and only present jumps in the angular momentum $L_z$ and Carter constant $Q$.

Reference~\cite{Gupta:2022jdt} derived (their Eq.~(26)) a relation between the jumps in $Q$ and $L_z$ for a stationary perturber, for resonances with $m\neq 0$. This is the so-called two-for-one deal. For $m=0$, axisymmetry enforces $\Delta L_z=0$ and therefore does not convey information about the $\Delta Q$, which  can be nonzero. We therefore compute $L_z$ and $Q$ independently and use the two-for-one deal whenever $m\neq 0$ as a cross-check.

Unless stated otherwise, we place the perturber at inclination $45^\circ$ relative to the equatorial plane along the $x$-axis. Other inclination angles follow from the analytic transformation law in Eq.~\ref{eq:transformation-with-inclination-angle-perturber} (App.~\ref{app:jumps}). In particular, for an equatorial perturber only even-$m$ resonances produce nonzero jumps in $L_z$ and $Q$, consistent with earlier work \cite{Gupta:2021cno}. Note that for all these amplitudes an overall scaling with $m_\star M^2/b^3$ is scaled out, see App.~\ref{app:jumps} again for more details.

In the remainder of this section, we present the jump-amplitude results (and the corresponding duration-weighted jumps) for our parameter choices, highlighting the dominant resonances and their dependence on the orbital configuration.

%%%%%%%%%%%%%%%%%%%%%%%%%%%%%%%%%%%%%%%%%%
\subsection{Prograde orbits}
%%%%%%%%%%%%%%%%%%%%%%%%%%%%%%%%%%%%%%%%%%

\begin{figure*}[h]
  \centering
  \begin{subfigure}{.32\textwidth}
      \includegraphics[width=1\textwidth]{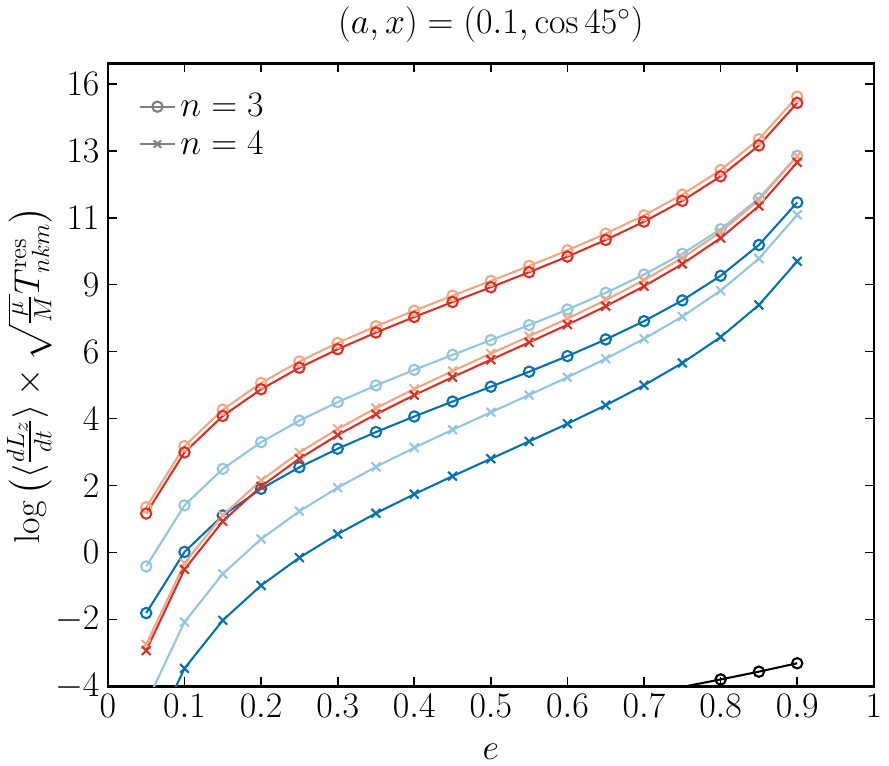}
      \centering
      \caption{}
      \label{fig:DurationxJump2D_L_a01_x07}
  \end{subfigure}
  \begin{subfigure}{.32\textwidth}
      \includegraphics[width=1\textwidth]{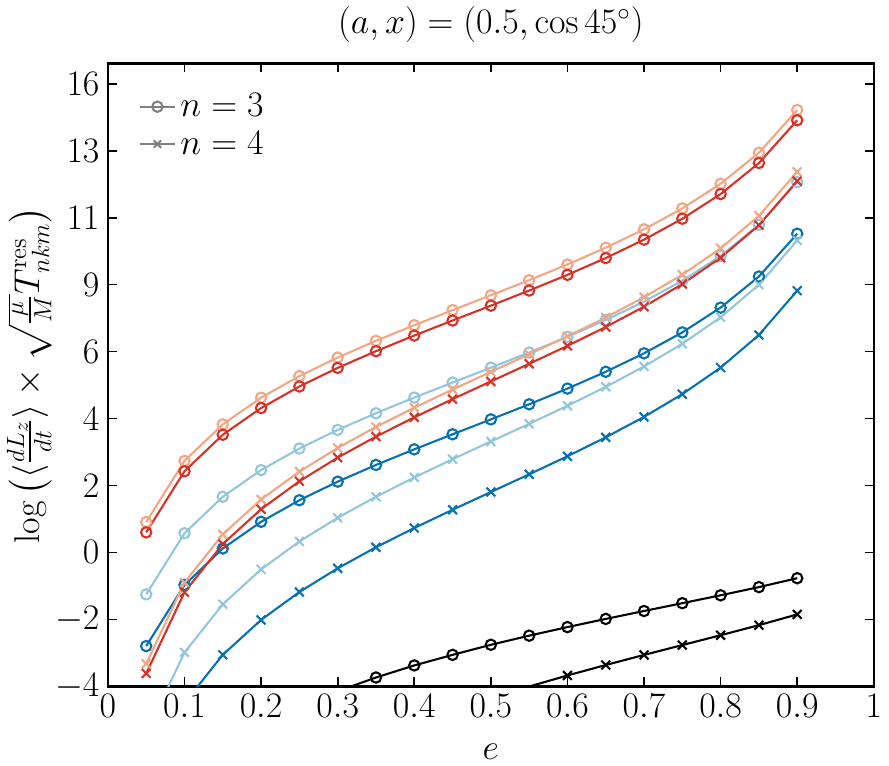}
      \centering
      \caption{}
      \label{fig:DurationxJump2D_L_a05_x07}
    \end{subfigure}
    \begin{subfigure}{.32\textwidth}
      \includegraphics[width=1\textwidth]{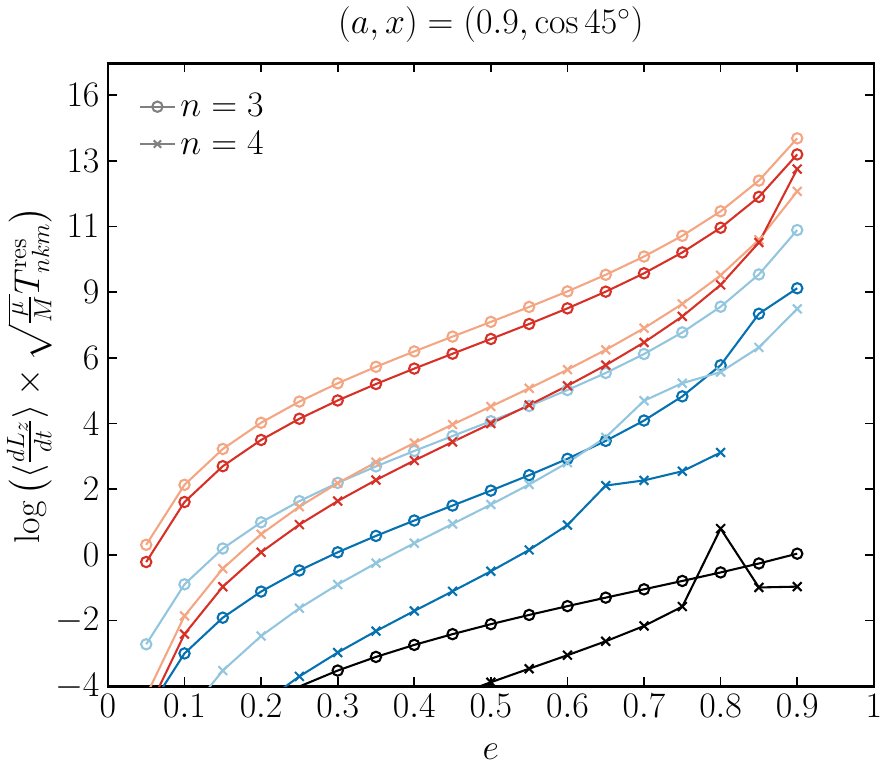}
      \centering
      \caption{}
      \label{fig:DurationxJump2D_L_a09_x07}
    \end{subfigure}
  \caption{Jump amplitudes for $L_z$ times resonance duration, as a function of eccentricity, for fixed spin $a$ and (prograde) inclination for all resonances with $n=3$ (circled) and $n=4$ (crossed). In both cases, the resonances with $k=0,\cdots,-4$ are colored in orange, red, black, light blue, and dark blue, respectively.We note that the angular-momentum jumps for the case $m=0$ ($k=-2$, black) vanish identically in the analytic treatment; nevertheless, they appear in the figure. This behavior arises because, in the numerical evaluation, the corresponding jump values are not exactly zero (although extremely small), and multiplication by the comparatively long resonance duration produces a finite contribution. We retain these curves to provide a qualitative indication of the numerical precision achieved.}
  \label{fig:DurationxJump2D_L_x07}
\end{figure*}

In Fig.~\ref{fig:DurationxJump2D_L_x07}, we show the `` resonance duration x jump amplitudes'' in the angular momentum $L_z$ for all the resonances, at fixed inclination $x = \cos 45^\circ$, with spin $a=0.1$ (\ref{fig:DurationxJump2D_L_a01_x07}), $a=0.5$ (\ref{fig:DurationxJump2D_L_a05_x07}), and $a=0.9$ (\ref{fig:DurationxJump2D_L_a09_x07}).
The main features are the rapid increase in the duration x jump at high $e$, and the vanishing of the jumps when $e \rightarrow 0$. Because the values are represented on a logarithmic scale, the values start at $e=0.05$.
The large $e$ behavior is due to the fact that the jumps diverge in the limit $e \to 1$. A possible explanation for this is that the secondary has a larger apoapsis and a smaller periapsis for such orbits. 
Meaning that $\mu$ comes closer to the perturber and the central BH at higher eccentricities, causing a stronger resonance and thus larger jump amplitudes. 
We do not have a rigorous explanation as to why the jumps vanish for circular orbits. 
Note that for the resonance $(3,-2,0)$, the associated jump in $L_z$ vanishes due to symmetry. Because the tidal force is symmetric in the $\phi$-direction when $m = 0$, there is no jump in $L_z$ for $m = 0$ resonances. The jump in $Q$ can be nonzero for these resonances, so these resonances are still relevant.
The numerical jump in $L_z$ for $(3,-2,0)$ can therefore be used as a rough measuring stick for the accuracy of the numerics. In particular, note that the numerical results become increasingly inaccurate with increasing $e$, and at large $a$, which is expected.
One thing to note is that the general and relative trends of each jump are qualitatively similar across the three values of $a$ we show, particularly for $a=0.1$ and $a=0.5$. Qualitatively, the jumps' amplitude decrease with increasing $a$.
In particular, we can create a rough hierarchy regarding the relative importance of each resonance: the two largest resonances, or roughly equal magnitudes correspond to $(3,0,-2)$ and $(3,-1,-1)$. This is then followed by nearly two orders of magnitudes lower by $(3,-3,1)$. The last $n=3$ resonance, namely $(3,-4,2)$ is typically about another two orders of magnitude smaller than the $(3,-3,1)$. Finally, the two largest jumps for the $n=4$ are the $(4,0,-2)$ and $(4,-1,-1)$ as well. Their importance is smaller than the $(3,-3,1)$ for low eccentricity, but become comparable to it at large eccentricity.
From the point of view of $L_z$ then, we can make a tier system:
\begin{enumerate}[noitemsep]
    \item Most important: $(3,0,-2)$, $(3,-1,-1)$.
    \item Less important: $(3,-3,1)$, and for higher eccentricity also $(4,-1,-1)$ and $(4,0,-2)$,
    \item Negligible: all the other.
\end{enumerate}

\begin{figure*}
  \centering
  \begin{subfigure}{.32\textwidth}
      \includegraphics[width=1\textwidth]{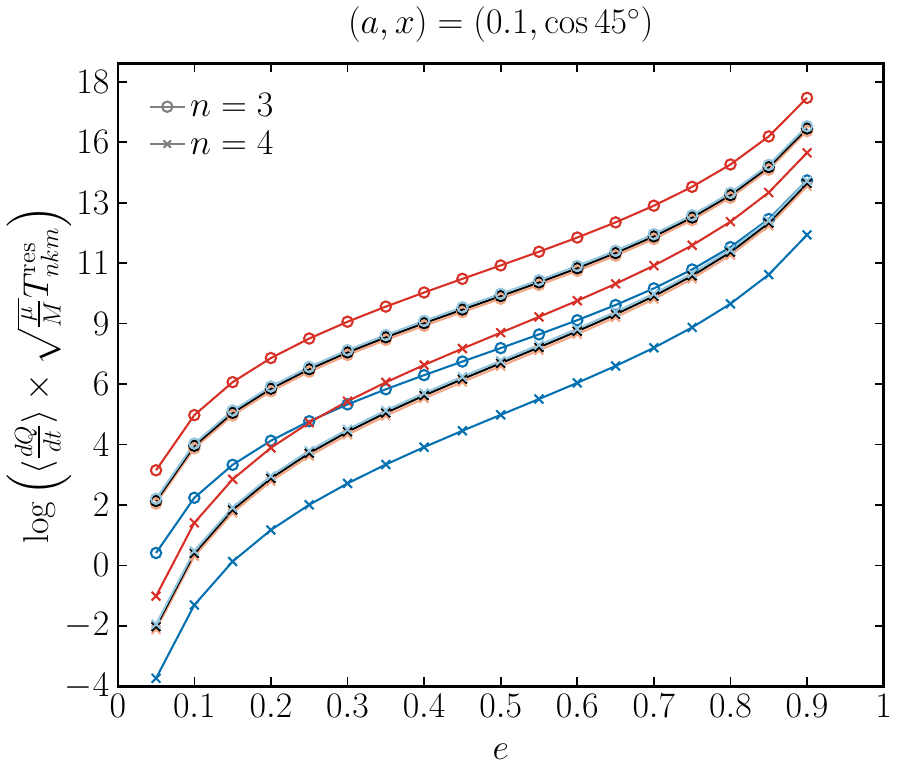}
      \centering
      \caption{}
      \label{fig:DurationxJump2D_Q_a01_x07}
  \end{subfigure}
  \begin{subfigure}{.32\textwidth}
      \includegraphics[width=1\textwidth]{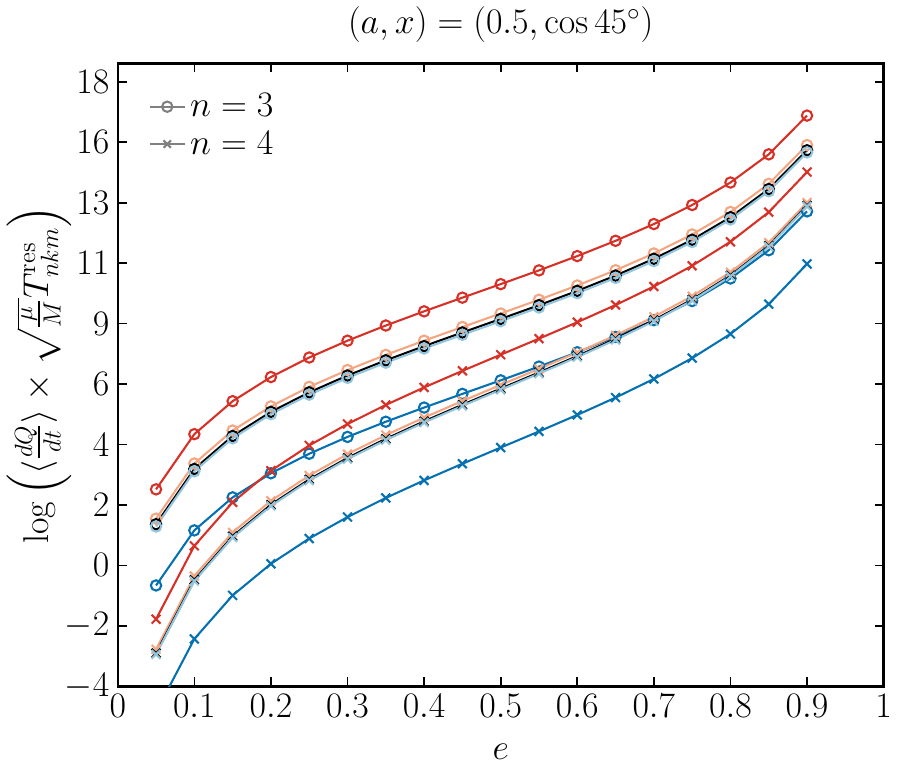}
      \centering
      \caption{}
      \label{fig:DurationxJump2D_Q_a05_x07}
    \end{subfigure}
    \begin{subfigure}{.32\textwidth}
      \includegraphics[width=1\textwidth]{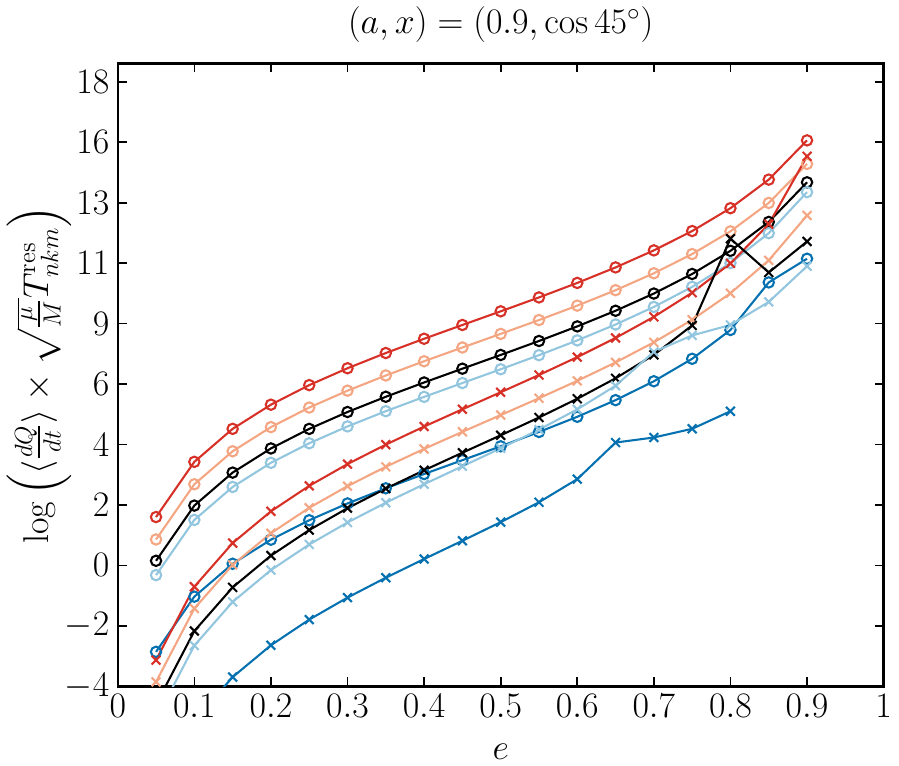}
      \centering
      \caption{}
      \label{fig:DurationxJump2D_Q_a09_x07}
    \end{subfigure}
  \caption{Jump amplitudes for $Q$ times resonance duration, as a function of eccentricity, for fixed spin $a$ and (prograde) inclination for all resonances with $n=3$ (circled) and $n=4$ (crossed). In both cases, the resonances with $k=0,\cdots,-4$ are colored in orange, red, black, light blue, and dark blue, respectively. }
  \label{fig:DurationxJump2D_Q}
\end{figure*}

In Fig.~\ref{fig:DurationxJump2D_Q}, we show the same plot as Fig.~\ref{fig:DurationxJump2D_L_x07}, but for the jump in $Q$. 

Much of the qualitative features are the same as for the jumps in $L_z$, with the exception that the resonances with $m=0$ do not vanish.
For $Q$ we can then also create a hierarchy:
\begin{enumerate}[noitemsep]
    \item Most important: $(3,-1,-1)$.
    \item Less important: $(3,0,-2)$. For lower spin values, $(3,-2,0)$ and $(3,-3,1)$ are about equally important. At high eccentricity, $(4,-1,-1)$ is also in that tier.
    \item Negligible in all cases: all other resonances.\footnote{By negligible, we mean that the jumps are typically less than 1/1000 of the highest resonance jump.}
\end{enumerate}
The precise tiering differs from that for $L_z$ because the hierarchy is based on \emph{relative} jump sizes. For $L_z$, the $(3,0,-2)$ and $(3,-1,-1)$ resonances typically produce jumps of the same order of magnitude and are therefore grouped together. For $Q$, the $(3,0,-2)$ jump is generally smaller---often by roughly an order of magnitude compared to the dominant $(3,-1,-1)$ resonance---so it is placed in the lower tier. This ranking is intended as a heuristic guide rather than a sharp classification.

\begin{figure*}
  \centering
  \begin{subfigure}{.47\textwidth}
      \includegraphics[width=\textwidth]{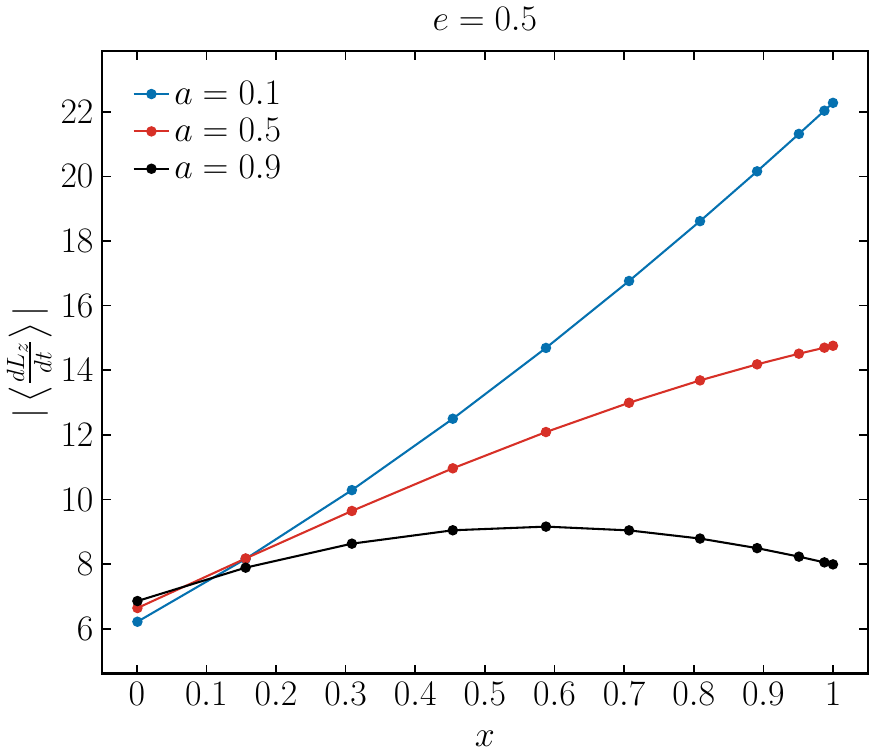}
      \centering
      \label{fig:ampse0.530-2}
  \end{subfigure}
  \begin{subfigure}{.47\textwidth}
      \includegraphics[width=\textwidth]{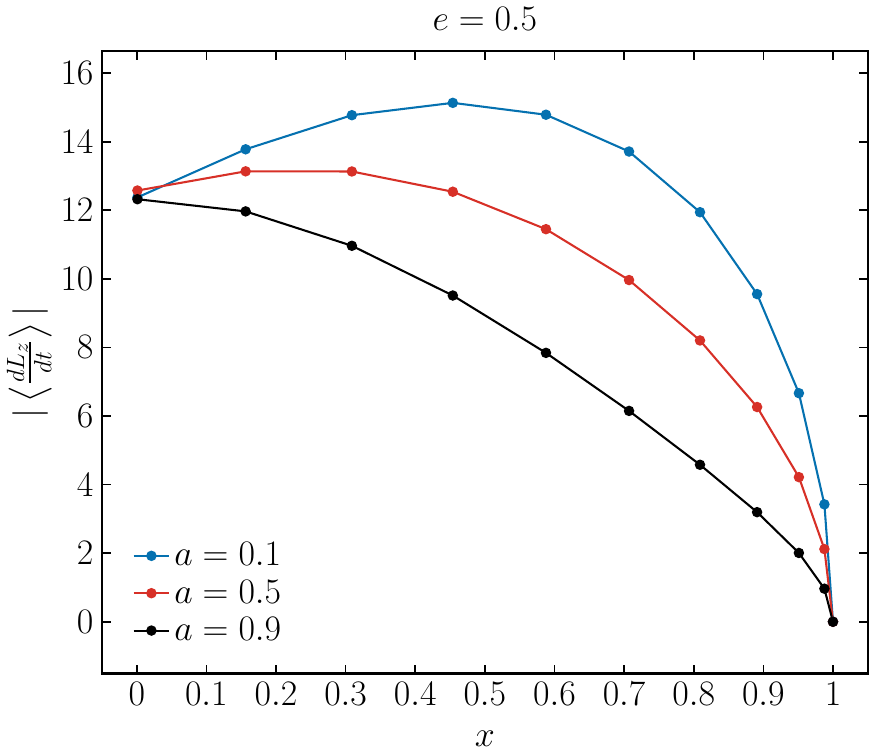}
      \centering
      \label{fig:ampse0.530-2}
  \end{subfigure}
  \begin{subfigure}{.47\textwidth}
      \includegraphics[width=\textwidth]{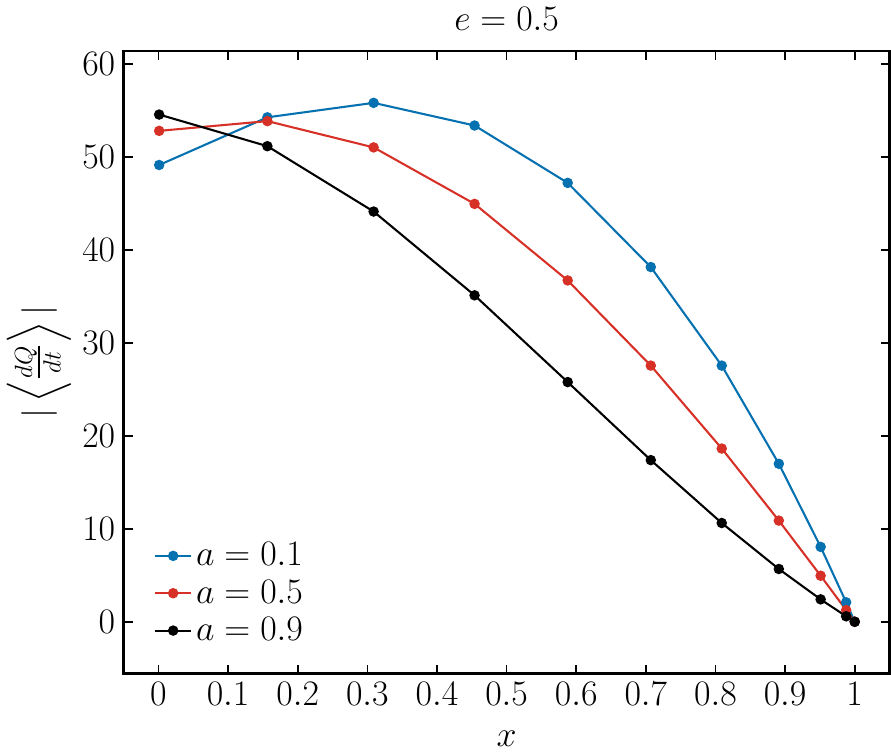}
      \centering
      \caption{$(n, k, m) = (3, 0, -2)$}
      \label{fig:ampse0.53-1-1}
      \end{subfigure}
\begin{subfigure}{.47\textwidth}
      \includegraphics[width=\textwidth]{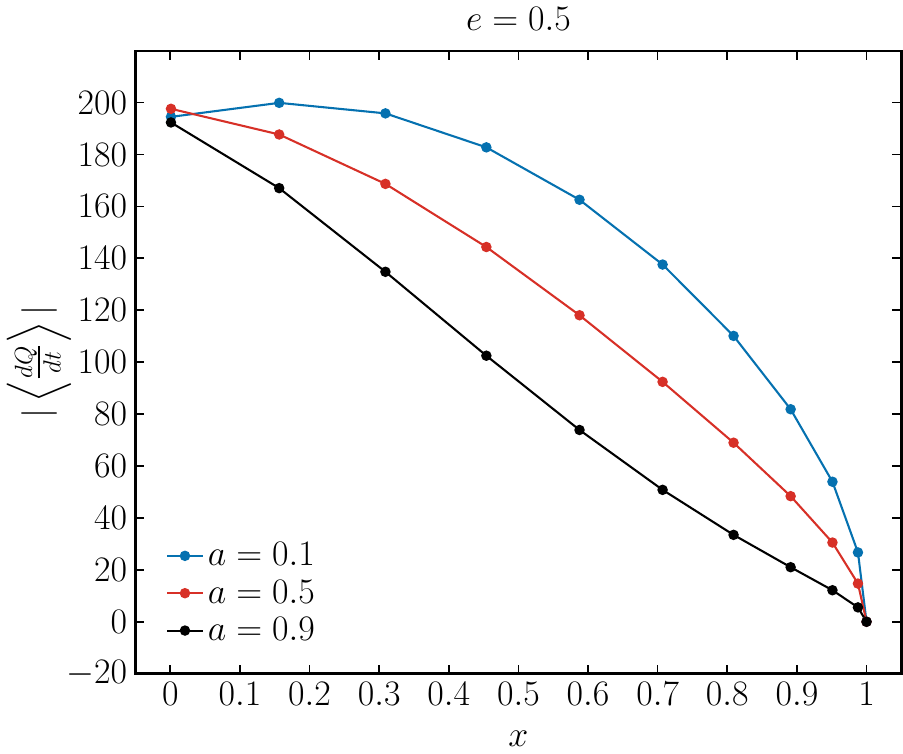}
      \centering
      \caption{$(n, k, m) = (3, -1, -1)$}
      \label{fig:ampse0.53-1-1}
      \end{subfigure}
  \caption{Imaginary part of the jump amplitudes in $L_z$ (bottom) and $Q$ (top) as function of $x$ for $e = 0.5$ and three values of $a$. 
  The plots on the left (a) belong to the ($3, 0, -2$) resonance, and the plots on the right (b) belong to the ($3, -1, -1$) resonance.}
  \label{fig:ampse0.5}
\end{figure*}

Figure \ref{fig:ampse0.5} shows the jump amplitudes as a function of inclination. The shape of the $x$ dependence of the jump amplitudes contains no clear patterns apart from the low $x$ and high $x$ features.
In particular, as $x \rightarrow 0$ (high inclination orbits), the dots representing the amplitudes at different $a$ values seem to move toward each other. 
This behavior is expected as spin effects should become smaller for orbits with a high inclination angle. 
The jump amplitudes belonging to the $(3,-1,-1)$ resonance seem to almost overlap at $I = 89^\circ$. 
This is different for the $(3,0,-2)$ resonance. At very low $x$, the jump amplitudes in $Q$ become sensitive again to the spin value.
The influence of the spin parameter at low $x$ seems to be dependent mostly on $k$ and $m$. The nature of this dependence is not yet understood. 
A deeper investigation on these jump amplitudes needs to be performed to provide details on this pattern.  
In the opposite limit, as $x \to 1$, the jump amplitudes in $Q$ vanish. This behavior is anticipated and follows from the functional dependence of the rate of change of $Q$ given in Eq.~\eqref{eq:dQdtau}.
For the $(3,-1,-1)$ resonance, the jump amplitudes in $L_z$ also approach zero as $x \to 1$, but this is not the case for the $(3,0,-2)$ resonance.
This difference can be understood in terms of the two-for-one deal \cite{Gupta:2022jdt}. As discussed earlier, that relation directly ties the jump amplitudes in $L_z$ to those in $Q$ whenever $m \neq 0$, with a simple proportionality constant.
Consequently, if the jump amplitude in $Q$ vanishes, the jump amplitude in $L_z$ must vanish as well, unless the proportionality constant itself is zero. One can show that this constant can vanish only if $k = 0$. In particular, for the $(3,0,-2)$ resonance we find that this proportionality constant does, in fact, vanish exactly.\footnote{In this case, $y_- = 0$ in the notation of \cite{Gupta:2022jdt}, which leads to an exact cancellation of the elliptic integrals in the two-for-one deal.} As a consequence, $L_z$ is left unconstrained. By contrast, for the $(3,-1,-1)$ resonance the proportionality constant is nonzero, and so the jump in $L_z$ also vanishes in the equatorial limit.

Finally, we note a general trend: the products ``resonance duration $\times$ jump amplitude'' are smallest at high spin $a$ and largest at low spin $a$. This behavior holds across all resonances with $n=3,4$ and reflects two comparable effects: for smaller $a$ the jump amplitudes are typically larger, and the resonances also tend to last longer.

%\begin{figure*}
%  \centering
%  \begin{subfigure}{.49\textwidth}
%      \includegraphics[width=1\textwidth]{Figures/n3k0m-2I50progamps.pdf}
%      \centering
%      \caption{$(n, k, m) = (3, 0, -2)$}
%      \label{fig:ampsI5030-2}
%\end{subfigure}
%  \begin{subfigure}{.50\textwidth}
%      \includegraphics[width=1\textwidth]{Figures/%n3k-1m-1I50progamps.pdf}
%      \centering
%      \caption{$(n, k, m) = (3, -1, -1)$}
%      \label{fig:ampsI503-1-1}
%      \end{subfigure}
%  \caption{Imaginary part of the jump amplitudes in $L_z$ (bottom) and %$Q$ (top) as function of $e$ for $I = 50^\circ$ and three values of $a$. 
%  The plots on the left (a) belong to the ($3, 0, -2$) resonance, and %the plots on the right (b) belong to the ($3, -1, -1$) resonance.}
%  \label{fig:ampsI50}
%\end{figure*}

\subsection{Retrograde orbits}

For completeness, we provide the same plots as we did for the prograde orbits in App.~\ref{app:plots}, Fig.~\ref{fig:DurationxJump2D_L_x-07}, and Fig.~\ref{fig:DurationxJump2D_Q_x-07}.

The general qualitative behavior of the jump amplitudes in the retrograde case are the same to the prograde case.
The only notable difference is that higher spin values now correspond to slightly higher jump amplitudes; the opposite behavior as for the prograde case.
A possible explanation of this trend is that, for prograde orbits, the resonance is closer to the central massive black hole at higher $a$, meaning that the resonance is weaker. 
For retrograde orbits the opposite happens. Higher values of the central BH spin result in the resonance occurring further away and thus a stronger resonance.

As we did for the prograde case, we can establish a rough hierarchy of most to least important resonances.
For jumps in $L_z$, we find:
\begin{enumerate}[noitemsep]
    \item Most important: $(3,-1,1)$ and $(3,0,2)$. 
    \item Less important: $(3,-3,-1)$, and for high spin $(4,-1,1)$ and $(4,0,2)$.
    \item Negligible: everything else.
\end{enumerate}
Finally, for the jumps in $Q$, we find
\begin{enumerate}[noitemsep]
    \item Most important: $(3,-1,1)$
    \item Less important:  $(3,0,2)$, $(3,-2,0)$, and $(3,-3,-1)$, and for high spin $(4,-1,1)$. 
    \item Negligible: everything else.
\end{enumerate}

\section{Discussion}
\label{sec:discussion}

Our results show that, in the stationary-perturber approximation considered here, a small subset of low-order modes dominates the 
phase-independent impact measure ``(jump amplitude)$\times$(resonance duration)'' over a broad region of parameter space.
For prograde orbits the most important resonances are typically $(3,-1,-1)$ and $(3,0,-2)$, with $(3,-3,1)$, $(4,-1,-1)$, and $(4,0,-2)$ forming the next tier. For retrograde orbits the dominant modes shift to $(3,-1,1)$ and $(3,0,2)$, followed (again) by $(3,-3,1)$ and the $n=4$ counterparts $(4,-1,1)$ and $(4,0,2)$. These lists provide a practical starting point for waveform modeling, where one wants to capture the largest dephasings at minimal computational cost.

At the same time, the full resonant kick contains a factor $e^{i\chi}$, with $\chi$ set by the orbital phases at resonance crossing.
This implies an important caveat: even resonances with comparatively small duration-weighted jumps can be crucial indirectly, because they can shift the orbital phases accumulated between resonances and therefore change the phase with which the system enters a later, larger resonance.
A robust modeling strategy should therefore treat the ``dominant-resonance'' list as an efficiency guide, not as a strict truncation criterion.

Implementing these effects in FEW \cite{Chua:2020stf,Katz:2021yft,Speri:2023jte,Chapman-Bird:2025xtd} requires care in two places. First, the inspiral integrator must reliably detect crossings of resonance contours; in practice this suggests adaptive stepping and an event-detection/root-finding layer for the resonance condition, rather than relying on a fixed step size (to avoid stepping over narrow or closely-spaced contours). Second, in regions where resonance curves come close, cross, or overlap, one must consistently apply multiple kicks---either sequentially with controlled ordering and error estimates, or through a local treatment that resolves the combined forcing. 

Finally, relaxing the stationarity assumption introduces a qualitatively new structure. If the perturber moves (for simplicity, consider a circular orbit), the resonance condition involves the perturber's azimuthal frequency $\Omega_{\phi,td}$ and an additional harmonic index $s$.
Because $\Omega_{\phi,td}$ depends only on $(M,m_\star,b)$ (and not on the EMRI parameters), it effectively shifts the corresponding $\omega_{nkms}$ curves and can produce more than one resonance for fixed $(n,k,m,s)$: $\omega_{nkms}$ may cross zero at low $p$ and again at higher $p$ after turning over.
For this reason, we expect possibly two resonances per $(n,k,m,s)$ in the dynamic case. This is distinct from the stationary case for which $(n,k,m)$ uniquely specify the resonance.
In addition to this new feature, the tidal forcing term is significantly more complex and a consistent implementation would also require a careful tracking of the orbital phase of the perturber. This is currently being investigated in a setting for which the dynamics of the perturbing object can be treated parametrically \cite{inpreparation}.

\clearpage
\section*{Acknowledgments}
We would like thank Christian Chapman-Bird, Ward ten Haaf, Pim Kelderman and Philip Lynch for countless discussions on tidal resonances. BB would also like to thank Marta Cocco for a stimulating discussion about the $m=0$ modes in the metric perturbation during the Capra meeting in Brussels.
PB acknowledges support from the Dutch Research Council (NWO) with file number OCENW.M.21.119. 
This work makes use of the Black Hole Perturbation Toolkit \cite{BHPToolkit}.

\appendix
%%%%%%%%%%%%%%%%%%%%%%%%%%%%%%
%%%%%%%%%%%%%%%%%%%%%%%%%%%%%%
\section{Stationary jump amplitudes}
\label{app:jumps}
%%%%%%%%%%%%%%%%%%%%%%%%%%%%%%
%%%%%%%%%%%%%%%%%%%%%%%%%%%%%%
This appendix derives the jump amplitudes that characterize the change in orbital actions (or equivalently, the orbital constants of motion 
$(E,L_z,Q)$) when an EMRI passes through a tidal resonance. We work in the action-angle formalism, following the two-timescale/adiabatic treatment of the perturbed dynamics, to obtain a closed-form expression for the jump amplitude in terms of the resonance duration, phase, and a Fourier coefficient of the tidal force; this expression is evaluated numerically in the main body of the paper. For more details on this derivation, see the earlier work on tidal resonances in \cite{Bonga:2019ycj,Gupta:2021cno,Gupta:2022fbe}. What is new compared to that earlier work is an analytic formula for how the jump amplitudes change with the orientation of the tidal perturber. All numerical results for the jump amplitudes in this paper are for a perturber located along the $x$-axis at an inclination of $45^\circ$; Eq.~\eqref{eq:dJdlambda-rotated} gives the jump amplitudes for an arbitrary sky location of the perturber.

Within the action-angle formalism, the dynamics can be expressed in terms of the angles $q_i$ and their conjugate actions $J_i$. The actions represent integrals of motion and, in particular, can be mapped to the orbital constants of motion $E$, $L_z$, and $Q$.  
In an EMRI, the actions change only slowly over time. Consequently, their evolution can be treated perturbatively.  
If the perturber is effectively stationary over the timescale of a tidal resonance, the evolution of the action-angle variables is governed by  
\begin{align}
    \frac{d q_i}{d t} &= \omega_i(J_j) + \eta \, g_{i, sf} (q_r, q_\theta, J_j) + \epsilon \, g_{i, td} (q_r, q_\theta, q_\phi, J_j) \notag \\
    & \quad + \mathcal{O}(\eta^2, \epsilon_2, \eta \epsilon), \\
    \frac{d J_i}{d t} &= \eta \, G_{i, sf} (q_r, q_\theta, J_j) + \epsilon \, G_{i, td} (q_r, q_\theta, q_\phi, J_j) \notag \\
    & \quad + \mathcal{O}(\eta^2, \epsilon_2, \eta \epsilon).
\end{align}
Here, the subscript ``$sf$'' denotes the self-force contribution, which encodes corrections from the secondary's own gravitational field and scales with the EMRI mass ratio $\eta = \mu/M$. The subscript ``$td$'' identifies the tidal contribution, sourced by the external perturber's gravitational field and scales as $\epsilon = m_\star M^2 / b^3$.

Because we want to study tidal resonance effects in the EMRI system, we will only focus on the tidal forces and omit the subscript ``$td$''. At leading adiabatic order, the above equation can be simplified to
\begin{align}
    \frac{d q_i}{d t} &\approx \omega_i(J_j), \label{eq:dqdtau} \\
     \frac{d J_i}{d t} &\approx \epsilon \langle G_{i} (q_r, q_\theta, q_\phi, J_j) \rangle.
\end{align}
The brackets $\langle \cdots \rangle$ denote a long time orbit averaging:
\begin{align}
    &\langle f(q_r,q_\theta,q_\phi,J_j)  \rangle =  \notag \\
    & \lim_{T \to \infty} \frac{1}{T} \int_{-T}^T f(q_r(t),q_\theta(t),q_\phi(t),J_j) dt \, . \label{eq:orbit-averaging}
\end{align}
This integration is done by decomposing $G_i$ into a Fourier series,
\begin{equation}
    G_i(q_r, q_\theta, q_\phi, J_j) = \sum_{n,k,m} G_{i,nkm}(J_j) e^{i (n q_r + k q_\theta + m q_\phi)}.
\end{equation}
Using Eq.~\eqref{eq:orbit-averaging} and Eq.~\eqref{eq:dqdtau}, we then obtain
\begin{align}
    \langle G_i(q_r, q_\theta, q_\phi, J_j) \rangle & = \sum_{n,k,m} G_{i,nkm}(J_j) \times \notag \\
    & \lim_{T \to \infty} \frac{1}{T} \int_{-T}^T e^{i t(n \omega_r + k \omega_\theta + m \omega_\phi)} dt \; ,
\end{align}
where for simplicity the initial phases are set to zero, however, they can easily be reinstated when needed (see Eq.~\eqref{JumpFormula}).
The integral vanishes for all $n,k,m$, except when $n,k,m$ are all zero or when the frequencies become commensurate, i.e.,
\begin{equation}
    \omega_{nkm} := n \omega_r + k \omega_\theta + m \omega_\phi = 0.
\end{equation}
For the tidal force considered here, the coefficient corresponding to $n,k,m$ being all zero is very small and completely negligible. However, this is not the case when some of the integers $n,k,m$ are non-zero and a resonance occurs.

Assuming then that the orbit passes through a resonance $(n,k,m)$ at time $t = t_0$, the associated jump sizes read~\cite{Flanagan:2010cd}:
\begin{align}
    \Delta J_i &\approx \epsilon \int_{-\infty}^\infty \langle G_i(q_r, q_\theta, q_\phi, J_j) \rangle dt, \\
    &= \epsilon \sum_{j=\pm 1} \sqrt{\frac{2 \pi}{\eta |\Gamma|}}\exp \left( {\rm sgn}(j \Gamma) \frac{i \pi}{4} + i j \chi_s \right) G_{i,jn,jk,jm}(J_j) \label{JumpFormula}
\end{align}
where $\chi_s = n q_r(t_0) + k q_\theta(t_0) + m q_\phi(t_0)$ is the total phase at the time of resonance $t_0$, while $\Gamma = n \dot{\omega}_r(t_0) + k \dot{\omega}_\theta(t_0) + m \dot{\omega}_\phi(t_0)$ is the derivative of the orbital frequencies with respect to Boyer-Lindquist time at the time of resonance. A loose physical interpretation of the above result is that the change in the orbital actions is determined by:
\begin{equation}
    \Delta J= \text{resonance duration} \times e^{i \text{phase}} \times \text{jump amplitude}.
\end{equation}

In order to find the jump amplitudes, we write the evolution of the orbital constants in terms of the acceleration caused by the tidal perturber:
\begin{align}
    \frac{dL_{z}}{d\tau}& = a_\phi, \label{eq:dLdtau}\\
    \frac{dQ}{d\tau} &= 2 u_\theta a_\theta - 2 a^2 {\rm cos}^2 \theta u_t a_t + 2 {\rm cot}^2 \theta u_\phi a_\phi \label{eq:dQdtau},
\end{align}
with the acceleration determined by Eq.~\eqref{eq:acc} and $\tau$ is proper time.
Recall that the energy is approximately conserved due to the fact that the tidal field is considered stationary during the resonance timescale.

Although the variations of $L_z$ and $Q$ in Eq.~\eqref{eq:dLdtau}-\eqref{eq:dQdtau} are most naturally written with respect to proper time, for practical purposes, their variations are most easily computed when differentiated with respect to Mino time $\lambda$.

We can extract the Fourier coefficients appearing in Eq.~\eqref{JumpFormula} by phase space averaging over the 3-torus spanned by the angles. Specifically, choose a resonance $(n,k,m)$ and a set of orbital constants $J_i$ (or equivalently $(a,e,x)$), consistent with this resonance, then the Fourier coefficients are obtained by evaluating the integral,
\begin{align}
    \left. \left\langle \frac{dJ_i}{d\lambda} \right\rangle \right|_{nkm}& = \frac{1}{(2 \pi)^3} \int_0^{2 \pi} dq_r \int_0^{2 \pi} dq_\theta \int_0^{2 \pi} dq_\phi \notag \\
    &\qquad \frac{dJ_i}{d\lambda} e^{-i (n q_r + k q_\theta + m q_\phi)} \; ,\label{eq:dCdt}
\end{align}
where henceforth it is understood that $dJ_i/d\lambda$ are evaluated at the prescribed orbital parameters $(a,p,e,x)$.

Fortunately, the above Mino-time average is straightforwardly related to the Boyer-Lindquist time average $t$ via the formula~(Eq.(9.4) in Ref.~\cite{Drasco:2005is}):
\begin{equation}
    \left. \left\langle \frac{dJ_i}{dt} \right\rangle \right|_{nkm} = \left. \frac{1}{\Gamma_t}\left\langle \frac{dJ_i}{d\lambda} \right\rangle \right|_{nkm},
\end{equation}
where $\Gamma_t$ is the temporal Mino-time frequency.

Eq.~\eqref{eq:dCdt} can be rewritten in a complete Mino time formulation by introducing three separate Mino times $\lambda_i$ for $i \in {r,\theta,\phi}$, which are related to the angles $q_i$ as
\begin{equation}
    q_i = \frac{2\pi}{\Lambda_i} \lambda_i,
\end{equation}
where $\Lambda_i$ are the Mino-time periods. Eq.~\eqref{eq:dCdt} then reads:
\begin{align}
    \left. \left\langle \frac{dJ_i}{dt} \right \rangle \right|_{nkm}& = \frac{1}{\Gamma_t \Lambda_r \Lambda_\theta \Lambda_\phi} \int_0^{\Lambda_r} d\lambda_r \int_0^{\Lambda_\theta} d\lambda_\theta \int_0^{\Lambda_\phi} d\lambda_\phi  \notag \\
    & \quad \frac{dJ_i}{d\lambda} e^{-2 \pi i (n  \frac{\lambda_r}{\Lambda_r} + k \frac{\lambda_\theta}{\Lambda_\theta} + m \frac{\lambda_\phi}{\Lambda_\phi})}.
\end{align}

The integral over $\lambda_\phi$ can be done analytically, since the rates are known functions of $z = e^{i q_\phi}$ (owing to the fact that the tidal field is solely comprised of a quadrupolar field),
\begin{equation}
    \frac{dJ_i}{d\lambda} = \sum_{|\mu| \leq 2} A_{\mu} z^\mu,
    \label{eq:dJdlambdaExpansion}
\end{equation}
where $A_{\mu}$ are functions of $\lambda_r$ and $\lambda_\theta$.
As a result, the $\lambda_\phi$ integration simply yields
\begin{equation}
    \frac{1}{2 \pi} \int_0^{2 \pi} dq_\phi \frac{dJ_i}{d\lambda} e^{-imq_\phi} = A_{m}.
\end{equation}
We are then left with the two remaining integrations 
\begin{align}
    \left. \left\langle \frac{dJ_i}{dt} \right\rangle \right|_{nkm} &= \frac{1}{\Gamma_t \Lambda_r \Lambda_\theta} \int_0^{\Lambda_r} d\lambda_r \int_0^{\Lambda_\theta} d\lambda_\theta \notag\\
    & \qquad 
    A_{m} e^{-2\pi i (n \frac{\lambda_r}{\Lambda_r} +k \frac{\lambda_\theta}{\lambda_\theta})} \; .
\end{align}
The decomposition in Eq.~\eqref{eq:dJdlambdaExpansion} assumes the perturber is located on the $x$-axis. We can generalize the formula by rotating the perturber around the $y$- and $z$-axes with angles $\theta_y$ and $\theta_z$, respectively.
If we perform such a rotation, Eq.~\eqref{eq:dJdlambdaExpansion} becomes
\begin{align}
    \frac{dJ_i}{d\lambda}(\theta_y,\theta_z) &= A_{-2} z^{-2} e^{2 \theta_z} \cos^2 \theta_y \nonumber\\
    &+ A_{-1} z^{-1} e^{\theta_z} \sin(2 \theta_y) \nonumber \\
    &+\frac{1}{2} A_{0} (3 \cos(2\theta_y)-1) \nonumber \\
    &+ A_{1} z e^{-\theta_z} \sin(2\theta_y) \nonumber \\
    &+ A_{2} z^2 e^{-2\theta_z} \cos^2 \theta_y \nonumber \\
    &=: \sum_{|\mu| \leq 2} A_{\mu} z^\mu e^{-\mu \theta_z} f_\mu(\theta_y),
    \label{eq:dJdlambda-rotated}
\end{align}
where the $f_\mu(\theta_y)$ are given in Table~\ref{tab:fmu_thetay}. 

The jump amplitudes for a perturber located at a generic position are thus given by
\begin{equation}
    \left. \left\langle\frac{dJ_i}{dt} \right\rangle \right|_{nkm}(\theta_y,\theta_z) =  \left. \left\langle\frac{dJ_i}{dt} \right\rangle \right|_{nkm}(0,0)\, e^{-m \theta_z}\, f_m(\theta_y). \label{eq:dJ-dlambda-different-angles}
\end{equation}
In our computations, we fix $\theta_z=0$. Some care is required when selecting the reference angle $\theta_y$: for instance, if one chooses $\theta_y = 0$, then $f_{\pm 1}(0) = 0$, and the corresponding jump amplitudes for $m=\pm1$ are identically zero; see again Table~\ref{tab:fmu_thetay}. Consequently, throughout this paper we adopt $\theta_y = \pi/4$ for all quoted jump values.  

More generally, the jump amplitudes defined at two distinct reference orientations, $(\theta_y,\theta_z)$ and $(\tilde{\theta}_y,\tilde{\theta}_z)$, are connected by
\begin{align} 
    \left. \left\langle\frac{dJ_i}{dt} \right\rangle \right|_{nkm}(\tilde{\theta}_y,\tilde{\theta}_z) & = \left. \left\langle\frac{dJ_i}{dt} \right\rangle \right|_{nkm}(\theta_y,\theta_z) \times \notag  \\
    & \qquad e^{-m (\tilde{\theta}_z-\theta_z)} \frac{f_m(\tilde{\theta}_y)}{f_m(\theta_y)} \; . \label{eq:transformation-with-inclination-angle-perturber}
\end{align}
This relation follows immediately by forming the ratio of Eq.~\eqref{eq:dJ-dlambda-different-angles} evaluated at the two different orientations. A word of caution: this transformation cannot be used to extrapolate from a reference configuration with $\theta_y=0$ for $m=\pm1$, because in that case the amplitudes vanish identically and thus provide no information that can be rescaled.

\begin{table}[h]
\centering
\setlength{\tabcolsep}{18pt}
\renewcommand{\arraystretch}{1.2}
\begin{tabular}{cc}
\hline
\hline
$m$ & $f_m(\theta_y)$ \\
\hline
$-2$ & $\cos^2 \theta_y$ \\
$-1$ & $\sin 2\theta_y$ \\
$0$ & $\frac{1}{2} (3 \cos 2\theta_y - 1)$ \\
$1$ & $\sin 2\theta_y$ \\
$2$ & $\cos^2 \theta_y$ \\
\hline
\hline
\end{tabular}
\caption{$f_m(\theta_y)$ for all $|m|\leq2$, as given in Eq.~\eqref{eq:dJdlambda-rotated}.}
\label{tab:fmu_thetay}
\end{table}

%%%%%%%%%%%%%%%%%%%%%%%%%%%%%%
%%%%%%%%%%%%%%%%%%%%%%%%%%%%%%
\section{Rate of change of orbital frequencies} \label{app:omega-dot}
\begin{figure*}[h]
    \centering
    
    % --- ROW 1 ---
    \begin{subfigure}[b]{0.32\textwidth}
        \centering
        \includegraphics[width=\textwidth]{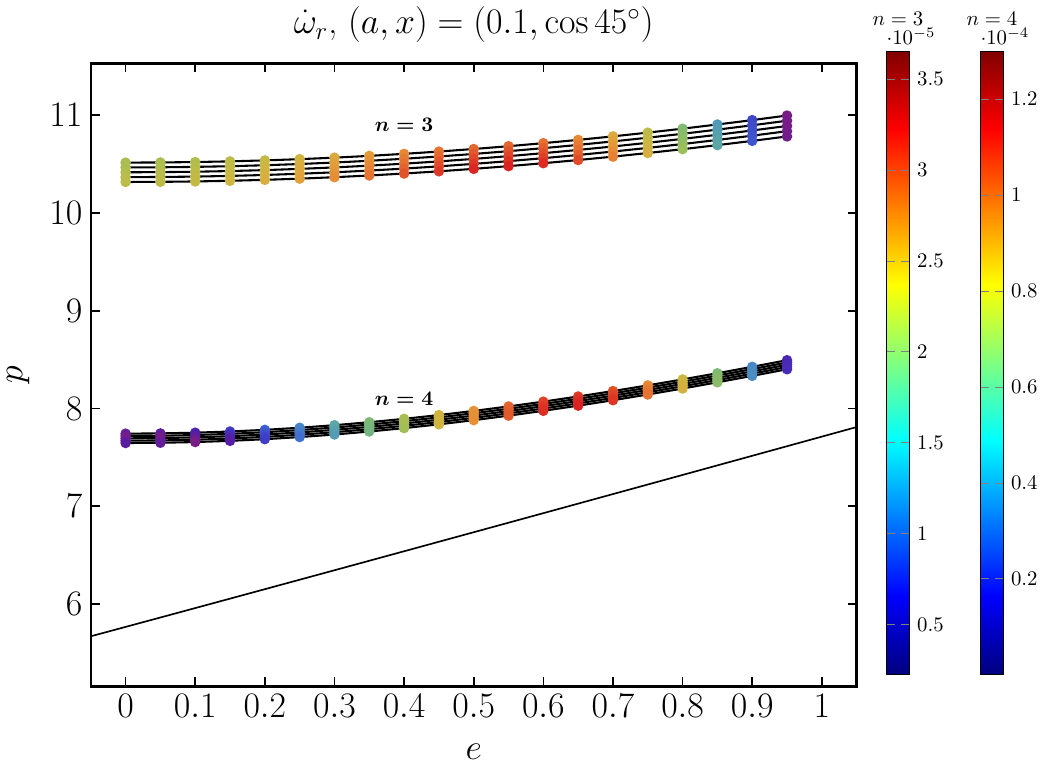}
        \caption{}
        \label{subfig:plot1}
    \end{subfigure}\hfill
    \begin{subfigure}[b]{0.32\textwidth}
        \centering
        \includegraphics[width=\textwidth]{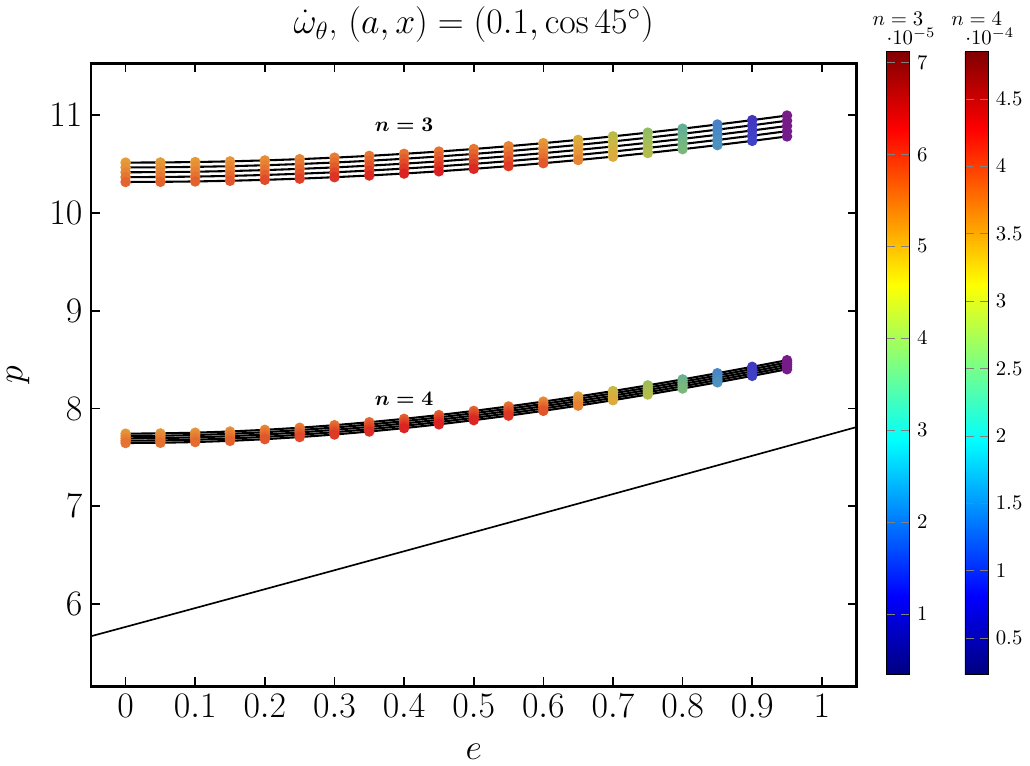}
        \caption{}
        \label{subfig:plot2}
    \end{subfigure}\hfill
    \begin{subfigure}[b]{0.32\textwidth}
        \centering
        \includegraphics[width=\textwidth]{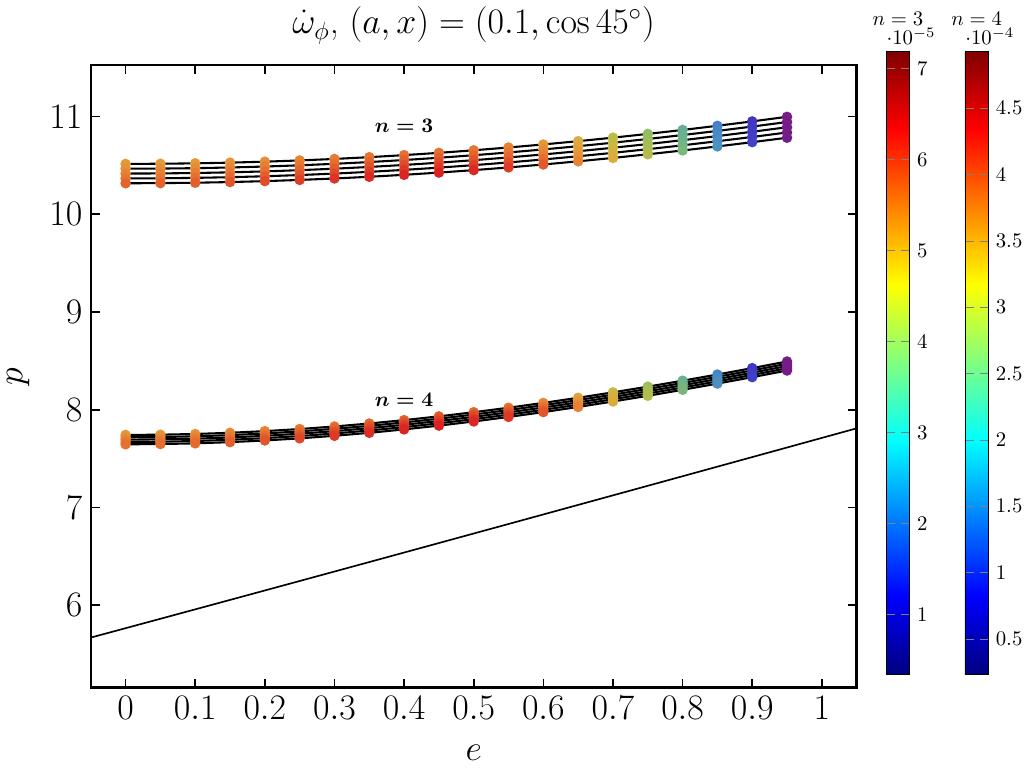}
        \caption{}
        \label{subfig:plot3}
    \end{subfigure}
    
    \vspace{0.5cm} % Vertical gap between rows
    
    % --- ROW 2 ---
    \begin{subfigure}[b]{0.32\textwidth}
        \centering
        \includegraphics[width=\textwidth]{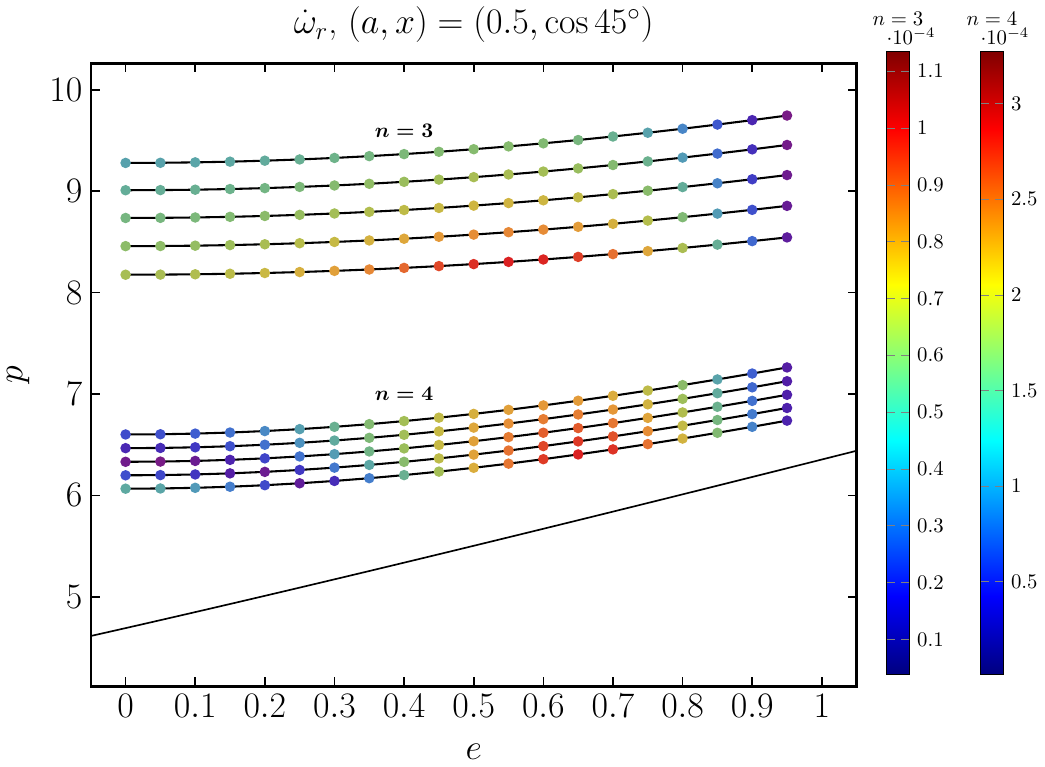}
        \caption{}
        \label{subfig:plot4}
    \end{subfigure}\hfill
    \begin{subfigure}[b]{0.32\textwidth}
        \centering
        \includegraphics[width=\textwidth]{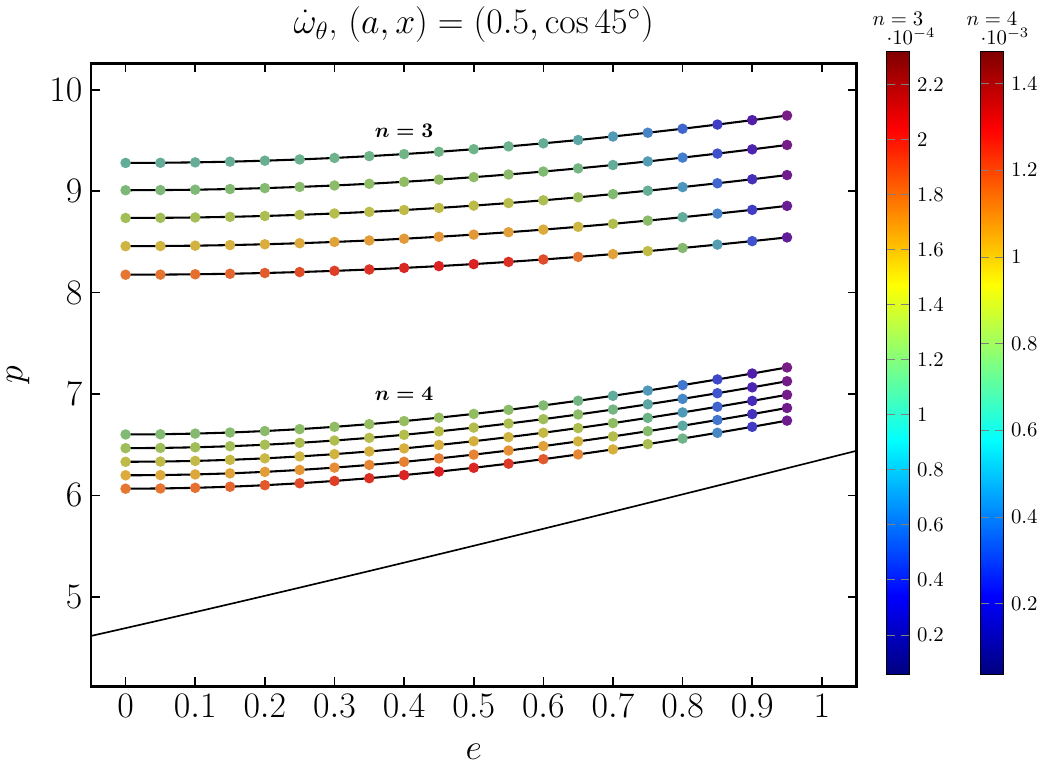}
        \caption{}
        \label{subfig:plot5}
    \end{subfigure}\hfill
    \begin{subfigure}[b]{0.32\textwidth}
        \centering
        \includegraphics[width=\textwidth]{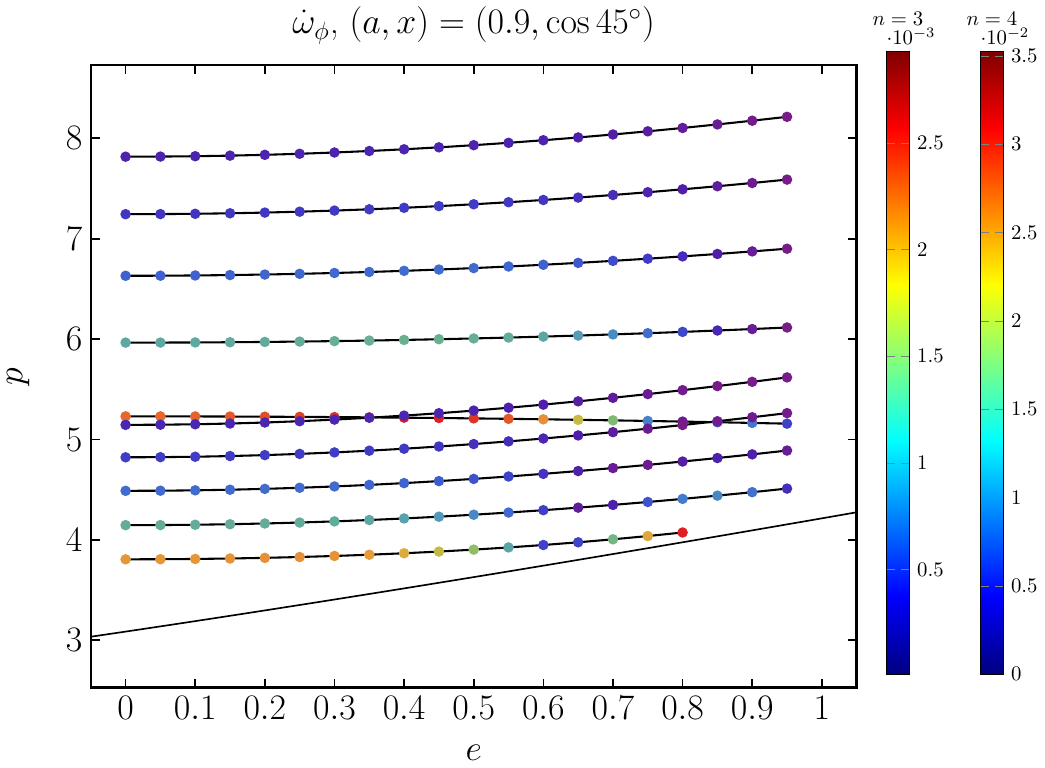}
        \caption{}
        \label{subfig:plot6}
    \end{subfigure}
    
    \vspace{0.5cm} % Vertical gap between rows
    
    % --- ROW 3 ---
    \begin{subfigure}[b]{0.32\textwidth}
        \centering
        \includegraphics[width=\textwidth]{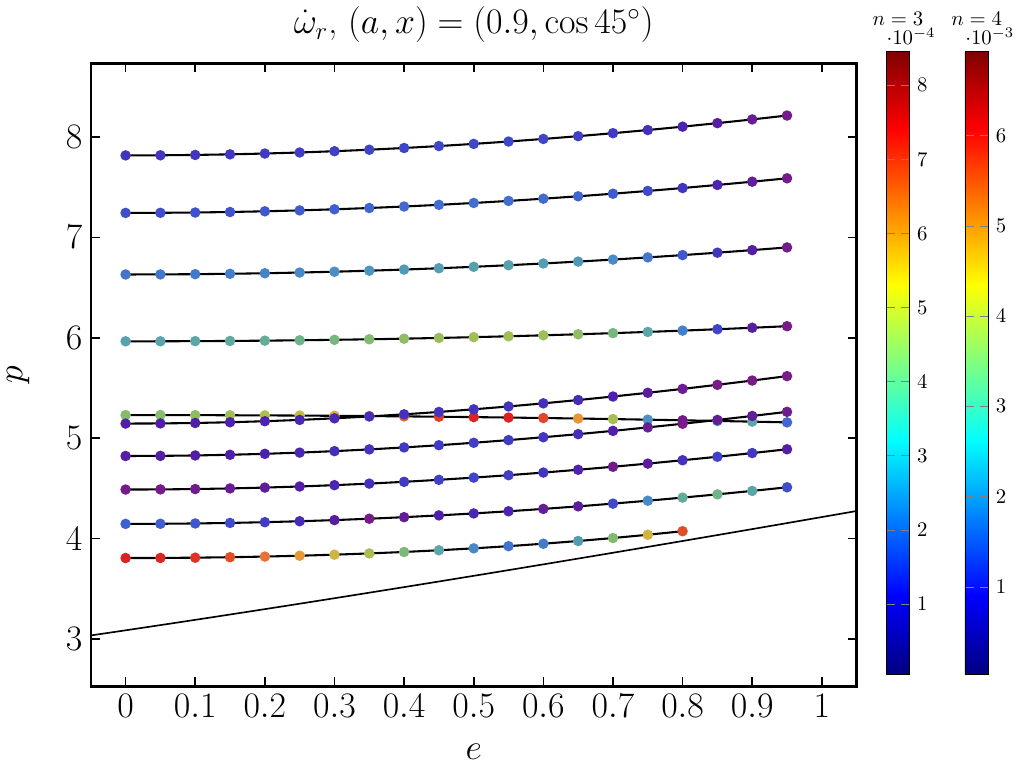}
        \caption{}
        \label{subfig:plot7}
    \end{subfigure}\hfill
    \begin{subfigure}[b]{0.32\textwidth}
        \centering
        \includegraphics[width=\textwidth]{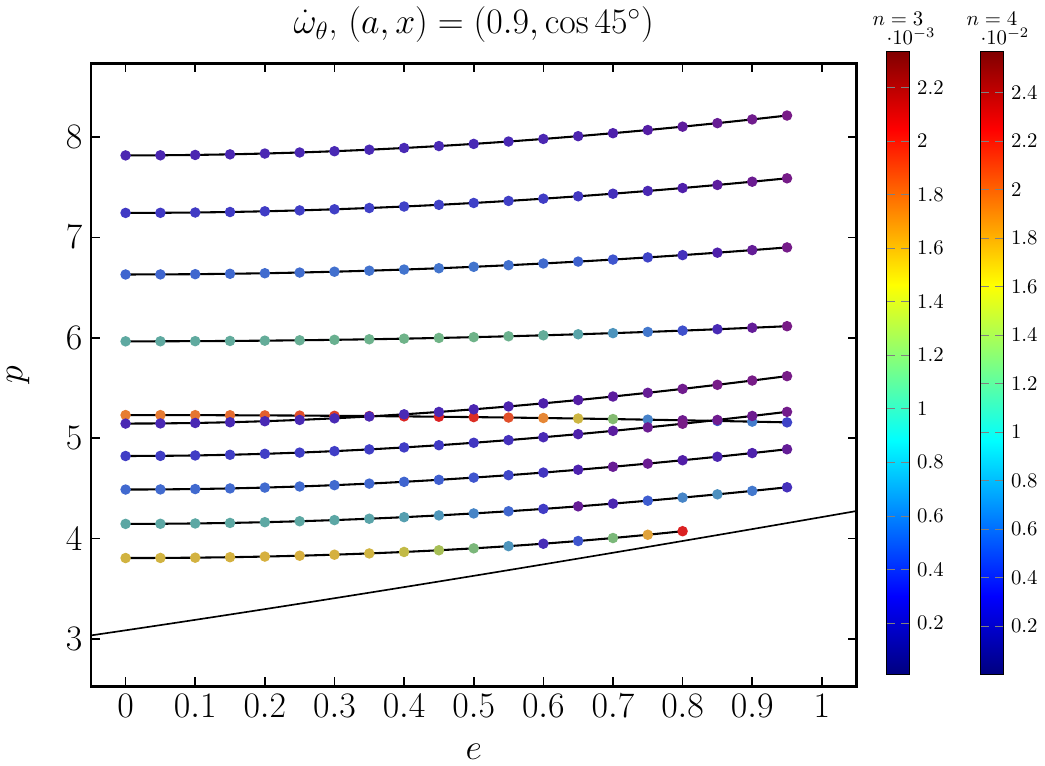}
        \caption{}
        \label{subfig:plot8}
    \end{subfigure}\hfill
    \begin{subfigure}[b]{0.32\textwidth}
        \centering
        \includegraphics[width=\textwidth]{OmegaDotphiColorMap_a09_x07}
        \caption{}
        \label{subfig:plot9}
    \end{subfigure}

    % Main overall caption for the entire 3x3 grid
    \caption{Heat map in $(p,e)$ of the individual derivative of the resonance frequencies, $\dot\omega_r$, $\dot\omega_\theta$, and $\dot\omega_\phi$, at fixed $(a,x)$. The color gradient correspond to the absolute value of $\dot\omega$. Resonance contours with the same $n$ share the same color gradient.}
    \label{fig:OmegaDotColorMap}
\end{figure*}

%%%%%%%%%%%%%%%%%%%%%%%%%%%%%%
In the main text, Fig.~\ref{fig:DurationColorMap} presents the resonance duration \eqref{eq:ResDuration} for all resonances as a heat map, computed using exact Kerr data for equatorial prograde orbits ($x=1$).
To identify the dominant factors determining the resonance duration, Fig.~\ref{fig:OmegaDotColorMap} displays a heat map of the derivatives of the orbital frequencies, $\dot\omega_r$, $\dot\omega_\theta$, and $\dot\omega_\phi$ with respect to Boyer-Lindquist time, evaluated along the resonance contours in the $(p,e)$ plane for fixed $(a,x)$. Since the durations cover comparable ranges at fixed $n$, we adopt a unified color scale for all resonances with the same $n$, where the scale is defined by the global minimum and maximum duration within that subset. The example shown corresponds to an inclined orbit, but the same qualitative behavior is observed for equatorial orbits.

%%%%%%%%%%%%%%%%%%%%%%%%%%%%%%

%%%%%%%%%%%%%%%%%%%%%%%%%%%%%%
%%%%%%%%%%%%%%%%%%%%%%%%%%%%%%
\section{Jumps for retrograde orbits}
\label{app:plots}
%%%%%%%%%%%%%%%%%%%%%%%%%%%%%%
%%%%%%%%%%%%%%%%%%%%%%%%%%%%%%

\begin{figure*}
  \centering
  \begin{subfigure}{.32\textwidth}
      \includegraphics[width=1\textwidth]{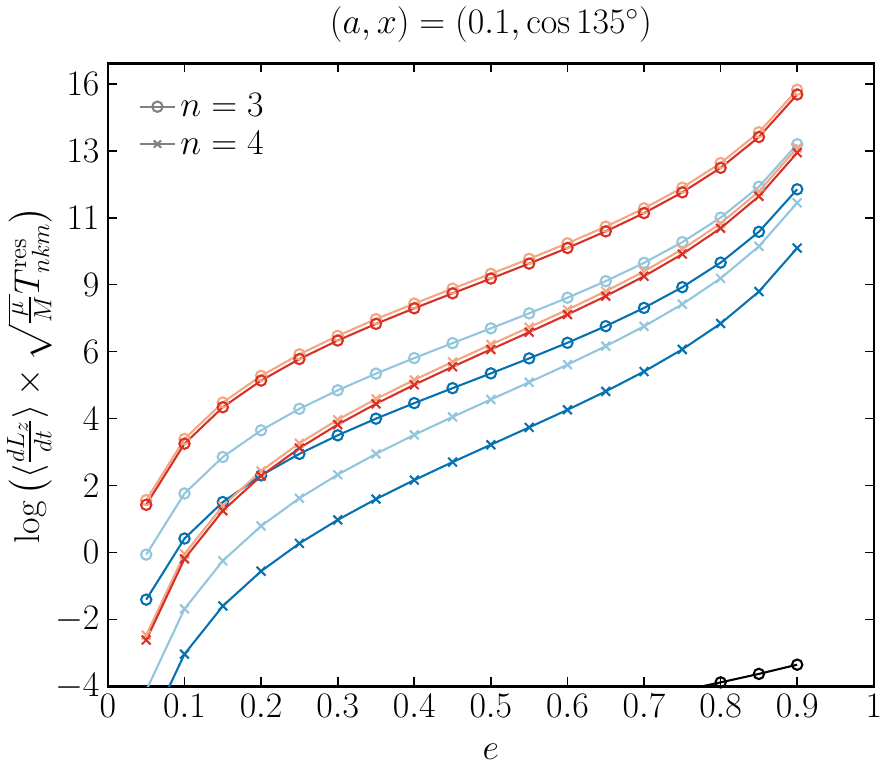}
      \centering
      \caption{}
      \label{fig:DurationxJump2D_L_a01_x-07}
  \end{subfigure}
  \begin{subfigure}{.32\textwidth}
      \includegraphics[width=1\textwidth]{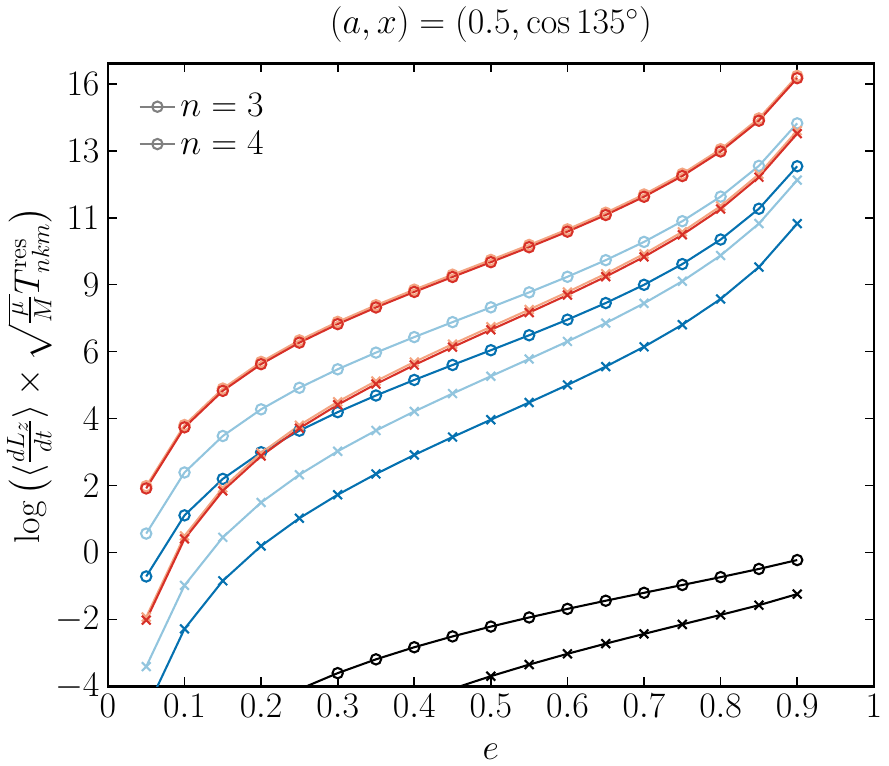}
      \centering
      \caption{}
      \label{fig:DurationxJump2D_L_a05_x-07}
    \end{subfigure}
    \begin{subfigure}{.32\textwidth}
      \includegraphics[width=1\textwidth]{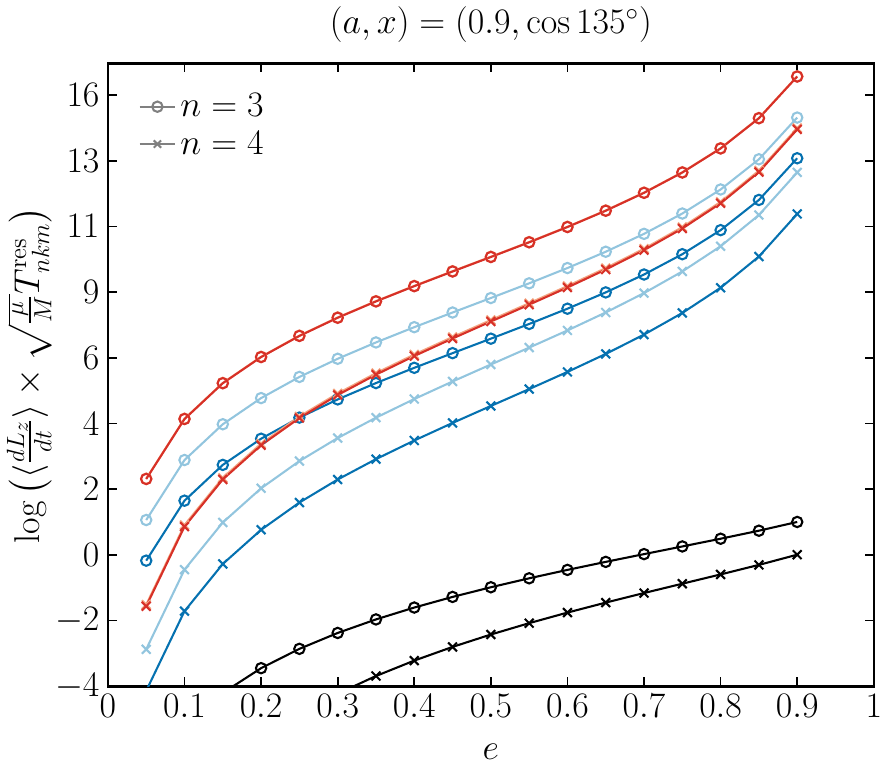}
      \centering
      \caption{}
      \label{fig:DurationxJump2D_L_a09_x-07}
    \end{subfigure}
  \caption{Jump amplitudes for $L_z$ times resonance duration, as a function of eccentricity, for fixed spin $a$ and (retrograde) inclination for all resonances with $n=3$ (circled) and $n=4$ (crossed). In both cases, the resonances with $k=0,\cdots,-4$ are colored in orange, red, black, light blue, and dark blue, respectively.}
  \label{fig:DurationxJump2D_L_x-07}
\end{figure*}

\begin{figure*}
  \centering
  \begin{subfigure}{.32\textwidth}
      \includegraphics[width=1\textwidth]{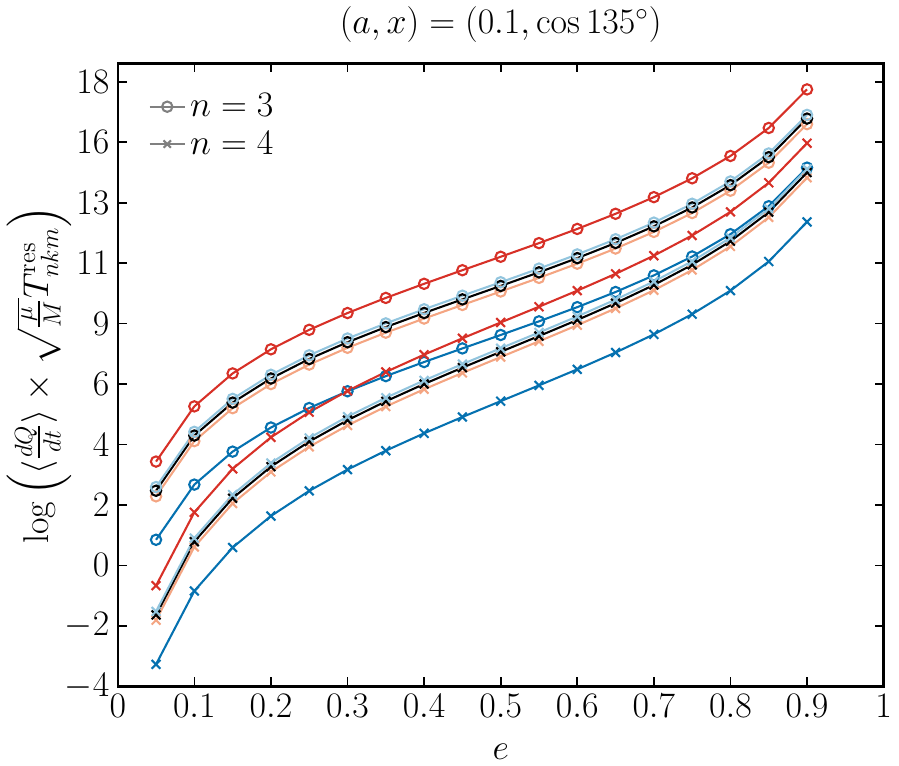}
      \centering
      \caption{}
      \label{fig:DurationxJump2D_Q_a01_x-07}
  \end{subfigure}
  \begin{subfigure}{.32\textwidth}
      \includegraphics[width=1\textwidth]{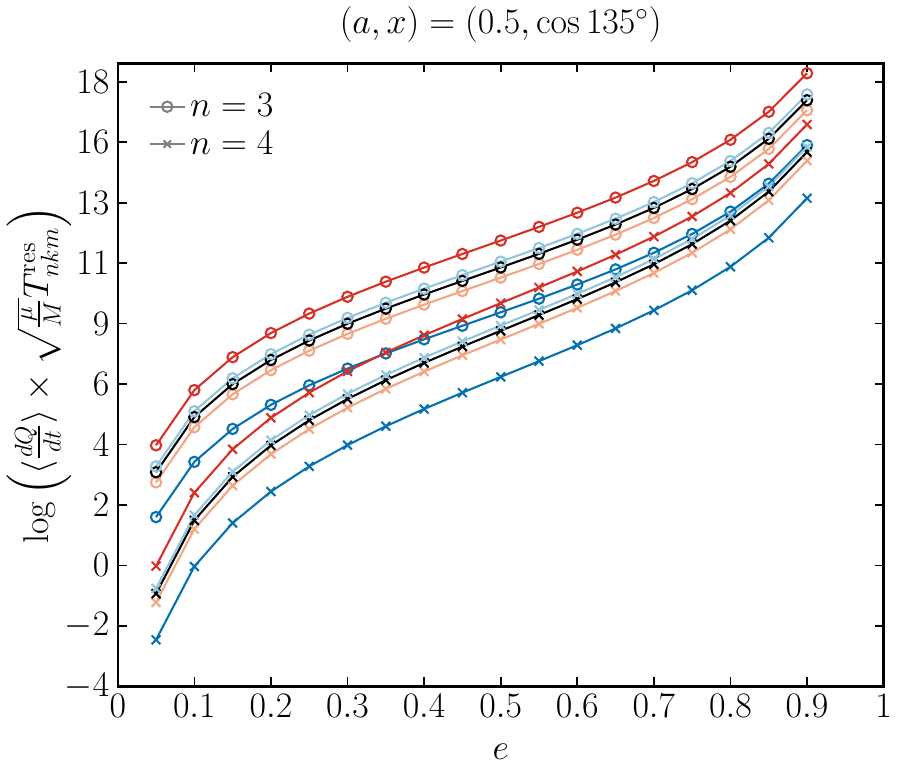}
      \centering
      \caption{}
      \label{fig:DurationxJump2D_Q_a05_x-07}
    \end{subfigure}
    \begin{subfigure}{.32\textwidth}
      \includegraphics[width=1\textwidth]{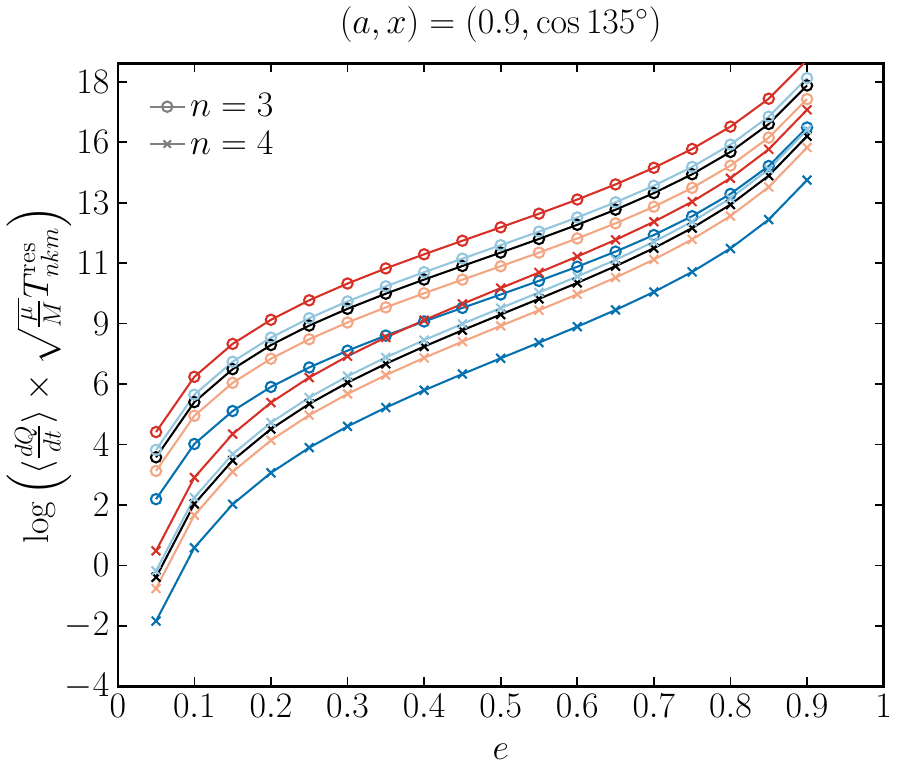}
      \centering
      \caption{}
      \label{fig:DurationxJump2D_Q_a09_x-07}
    \end{subfigure}
  \caption{Jump amplitudes for $Q$ times resonance duration, as a function of eccentricity, for fixed spin $a$ and (retrograde) inclination for all resonances with $n=3$ (circled) and $n=4$ (crossed). In both cases, the resonances with $k=0,\cdots,-4$ are colored in orange, red, black, light blue, and dark blue, respectively.} 
  \label{fig:DurationxJump2D_Q_x-07}
\end{figure*}

In Fig.~\ref{fig:DurationxJump2D_L_x-07} and Fig.~\ref{fig:DurationxJump2D_Q_x-07}, we show the product of the jumps amplitudes, $\langle \frac{dJ_i}{dt} \rangle$ with the resonance duration, $T^\text{res}_{nkm}$, for $J_i = L_z$ and $Q$, for retrograde orbits $x = \cos 135^\circ$, at different spins $a=0.1,0.5,0.9$.

%%%%%%%%%%%%%%%%%%%%%%%%%%%%%%
%%%%%%%%%%%%%%%%%%%%%%%%%%%%%%
\clearpage
\bibliography{bib-resonances}% Produces the bibliography via BibTeX.
\end{document}